%% file: 0_main.tex
\documentclass[runningheads]{llncs}

\input{0_packages}
\input{0_macros}
\input{0_plots}

\begin{document}

% Title brainstorming:
% 
% MDP Model-Checking Algorithms Revisited
% MDP Model-Checking Algorithms Revisited: Beyond Value Iteration
% MDP Model-Checking Algorithms Revisited: The Land Beyond Value Iteration
% MDP Model-Checking Algorithms Revisited, Experimentally
% Precision and Speed in MDP model checking
% An Experimental Comparison of MDP Model Checking Algorithms
% Algorithms for Fast and Precise MDP model checking
% A Practitioner's Guide to MDP Model Checking Algorithms
% Linear Programs, Value Iteration and Policy Iteration: The final showdown
% How to model check your MDP today?
% How do you want to model-check your MDP today?
% Linear Programs, Value Iteration or Policy Iteration: How do you want to model-check your MDP today?

\title{A Practitioner's Guide to\\MDP Model Checking Algorithms%
\thanks{This research was funded % in part by
by the European Union's Horizon 2020 research and innovation programme under the Marie Sk{\l}odowska-Curie grant agreement No.\ 101008233 (MISSION),
%by the EU's Horizon 2020 research and innovation programme under MSCA grant agreement 101008233 (MISSION), % shorter version
and by NWO VENI grant no.\ 639.021.754.%
}}
\titlerunning{A Practitioner's Guide to MDP Model Checking}
% If the paper title is too long for the running head, you can set
% an abbreviated paper title here
%
\author{%
Arnd~Hartmanns\inst{1}\,\orcidID{0000-0003-3268-8674} \and
Sebastian~Junges\inst{2}\,\orcidID{0000-0003-0978-8466} \and\\
Tim~Quatmann\inst{3}\,\orcidID{0000-0002-2843-5511} \and
Maximilian~Weininger\inst{4}\,\orcidID{0000-0002-0163-2152}
}

\authorrunning{A.\ Hartmanns et al.}
%\authorrunning{Authors omitted}
% First names are abbreviated in the running head.
% If there are more than two authors, 'et al.' is used.
%
\institute{
University of Twente, Enschede, The Netherlands
\email{a.hartmanns@utwente.nl} \and
Radboud University, Nijmegen, The Netherlands
\email{sebastian.junges@ru.nl} \and
RWTH Aachen University, Aachen, Germany
\email{tim.quatmann@cs.rwth-aachen.de} \and
Technical University of Munich, Munich, Germany
\email{maxi.weininger@tum.de} 
}
%\institute{}

\maketitle     

\vspace{-0.8cm}
\begin{abstract}
Model checking undiscounted reachability and expected-re\-ward properties on Markov decision processes (MDPs) is key for the verification of systems that act under uncertainty.
Popular algorithms are policy iteration and variants of value iteration;
in tool competitions, most participants rely on the latter.
These algorithms generally need worst-case exponential time.
However the problem can equally be formulated as a linear program, solvable in polynomial time.
In this paper, we give a detailed overview of today's state-of-the-art algorithms for MDP model checking with a focus on performance and correctness.
We highlight their fundamental differences, and describe various optimisations and implementation variants.
We experimentally compare floating-point and exact-arithmetic implementations of all algorithms on three benchmark sets using two probabilistic model checkers.
Our results show that (optimistic) value iteration is a sensible default, but other algorithms are preferable in specific settings.
This paper thereby provides a guide for MDP verification practitioners---tool builders and users alike.
%Thus, they are a relevant tool in the toolbox of the practitioner.
% \keywords{Quantitative model checking \and Markov decision process  \and linear programming \and value iteration \and more?}
\end{abstract}
\vspace{-1.3cm}

\input{1_introduction}
\input{2_prelims}
\input{3_zoomLP}

\input{4_zoomPI}
\input{5_qcomp}
\input{7_related}
\input{9_conclusion}

\clearpage

\bibliographystyle{splncs04}
\bibliography{ref}

\clearpage
\input{appendix.tex}

\end{document}

%% file: 0_packages.tex
\usepackage[T1]{fontenc}
\usepackage{graphicx}
\usepackage[bookmarks,unicode,colorlinks=true,breaklinks=true]{hyperref}
\usepackage{color}

\usepackage{xspace}
\usepackage[tableposition=top]{caption}
\usepackage{wrapfig}

\usepackage{amssymb}
\usepackage{mathtools}
\usepackage{microtype}

\usepackage{booktabs}
\usepackage{caption}
\usepackage{subfigure}
\usepackage{multicol}
\usepackage{xifthen}
\usepackage{etoolbox}
\usepackage{tikz}
\usetikzlibrary{arrows,shapes,calc,automata,positioning}
\tikzstyle{state} = [draw,fill=white,circle,thick,align=center,inner sep=0pt,minimum size=4.5mm]
\tikzstyle{lstate} = [draw,fill=white,rectangle,rounded corners,thick,align=center,inner sep=2pt]
\tikzstyle{dot} = [fill,circle,inner sep=0mm,minimum size=1.25mm,line width=0mm]

\usepackage{pgfplots}
\pgfplotsset{compat=newest}

\usetikzlibrary{shapes,positioning,arrows,calc,automata,matrix,fit}

\usepackage{xfrac}

\usepackage[capitalise]{cleveref}

\makeatletter%
\g@addto@macro\normalsize{%
  \setlength\abovedisplayskip{3pt}%
  \setlength\belowdisplayskip{3pt}%
  \setlength\abovedisplayshortskip{-3pt}%
  \setlength\belowdisplayshortskip{3pt}%
}%
\makeatother

%% file: 0_macros.tex
\renewcommand{\paragraph}[1]{\smallskip\noindent\emph{#1}}

\newcommand{\mdp}{\mathcal{M}}
\newcommand{\states}{\mathsf{S}}
\newcommand{\actions}{\mathsf{A}}
\newcommand{\trans}{\delta}

\newcommand{\dist}{\mathsf{d}}
\newcommand{\dists}{\mathsf{D}}
\newcommand{\eqdef}{\vcentcolon=}
\newcommand{\Rationals}{\mathbb{Q}}
\DeclareMathSymbol{\sm}{\mathbin}{AMSa}{"39} % short minus
\newcommand{\set}[1]{\ensuremath{\{\,#1\,\}}}
\newcommand{\tuple}[1]{\ensuremath{( #1 )}}

\newcommand{\apath}{\rho}
\newcommand{\policy}{\pi}
\newcommand{\policies}{\Pi}
\newcommand{\paths}{\mathsf{Paths}}
\newcommand{\prob}{\mathbb{P}}

\newcommand{\targets}{\mathsf{T}}
\newcommand{\opt}{\mathsf{opt}}
\renewcommand{\min}{\ensuremath{\mathrm{min}}}
\renewcommand{\max}{\ensuremath{\mathrm{max}}}
\newcommand{\val}{\mathsf{V}}
\newcommand{\rew}{\mathsf{rew}}
\newcommand{\Pobj}[2]{\ensuremath{\mathrm{P}_{{\!}#1}(#2)}}
\newcommand{\Eobj}[2]{\ensuremath{\mathrm{E}_{#1}(#2)}}

\newcommand{\Pmax}{\ensuremath{\mathrm{P}_{\max}}\xspace}

\newcommand{\lowerbound}{lb}
\newcommand{\upperbound}{ub}

\newcommand{\ie}{i.e.\xspace}
\newcommand{\eg}{e.g.\xspace}

\newcommand{\cf}{cf.\xspace}
\newcommand{\etal}{et al.\xspace}

\newcommand{\tool}[1]{\textsf{#1}\xspace}
\newcommand{\storm}{\tool{Storm}}
\newcommand{\mcsta}{\tool{mcsta}}
\newcommand{\solver}[1]{\textsf{#1}\xspace}

\makeatletter
\RequirePackage[bookmarks,unicode,colorlinks=true,breaklinks=true]{hyperref}%
   \def\@citecolor{blue}%
   \def\@urlcolor{blue}%
   \def\@linkcolor{blue}%

\def\orcidID#1{\smash{\href{http://orcid.org/#1}{\protect\raisebox{-1.25pt}{\protect\includegraphics{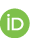}}}}}
\makeatother

%% file: 0_plots.tex
\newcommand{\numqvbsfull}{367}
\newcommand{\numqvbshard}{18}
\newcommand{\numpremise}{200}
\newtoggle{showplots}
\toggletrue{showplots}

\definecolor{plotred}{RGB}{255,0,0}
\definecolor{plotgreen}{RGB}{0,255,0}
\definecolor{plotblue}{RGB}{0,0,255}
\definecolor{plotyellow}{RGB}{230,230,0}
\definecolor{plotcyan}{RGB}{0,255,255}
\definecolor{plotorange}{RGB}{255,127,0}
\definecolor{plotpink}{RGB}{255,0,255}
\definecolor{plotlightgray}{RGB}{192,192,192}
\definecolor{plotdarkgray}{RGB}{128,128,128}
\definecolor{plotdarkred}{RGB}{128,0,0}
\definecolor{plotgreenyellow}{RGB}{128,128,0}
\definecolor{plotdarkgreen}{RGB}{0,128,0}
\definecolor{plotpurple}{RGB}{128,0,128}
\definecolor{plotteal}{RGB}{0,128,128}
\definecolor{plotdarkblue}{RGB}{0,0,128}
\definecolor{plotlightred}{RGB}{205,92,92}
\definecolor{plotlightblue}{RGB}{176,196,222}
\colorlet{color1}{plotred}
\colorlet{color2}{plotgreen}
\colorlet{color3}{plotblue}
\colorlet{color4}{plotyellow}
\colorlet{color5}{plotcyan}
\colorlet{color6}{plotorange}
\colorlet{color7}{plotpink}
\colorlet{color8}{plotlightgray}
\colorlet{color9}{plotdarkgray}
\colorlet{color10}{plotdarkred}
\colorlet{color11}{plotgreenyellow}
\colorlet{color12}{plotdarkgreen}
\colorlet{color13}{plotpurple}
\colorlet{color14}{plotteal}
\colorlet{color15}{plotdarkblue}
\colorlet{color16}{plotlightred}
\colorlet{color17}{plotlightblue}

\newcommand{\quantileplotxlabel}{}
\newcommand{\quantileplotylabel}{}
\newlength{\quantileplotwidth}
\newlength{\quantileplotheight}
\newcommand{\quantileplotlegendcols}{1}

\newcommand{\quantileplot}[8]{%
	\begin{tikzpicture}
	\begin{axis}[
	width=\quantileplotwidth,
	height=\quantileplotheight,
	xmin=#4,
	xmax=#5,
	ymin=#6,
	ymax=#7,
	ymajorgrids,
	ymode=log,
	axis x line=bottom,
	axis y line=left,
	unbounded coords=discard,filter discard warning=false, % properly deal with missing data points
	xlabel=\quantileplotxlabel,
	ylabel=\quantileplotylabel,
	log ticks with fixed point, % enable to avoid 10^-x notation
	yticklabel style={font=\scriptsize},
	scaled y ticks=false,
	xticklabel style={font=\scriptsize},
	legend columns=\quantileplotlegendcols,
	legend pos={#8},
	legend style={nodes={scale=0.75, transform shape},inner sep=1pt},
/pgfplots/legend image code/.code={\draw[mark repeat=2,mark phase=2,##1] plot coordinates {(0cm,0cm) (0.3cm,0cm)};}, % make the legend lines a bit smaller
	every axis plot/.append style={ultra thick},
	legend cell align={left}
	]
	\iftoggle{showplots}{
	\foreach \tool\color in {#2}{%
		\edef\loopbody{
			\noexpand\addplot[\color] table [x=n,y=\tool shifted, col sep=tab] {#1};
		}
		\loopbody
	}
	\legend{#3}}{\node[anchor=south west, align=center, red] {\huge NOT\\ COMPILED};}
	\end{axis}
	\end{tikzpicture}%
}

\newlength\scatterplotsize
\newcommand{\scatterplot}[6]{%
	\begin{tikzpicture}
	\begin{axis}[
	width=\scatterplotsize,
	height=\scatterplotsize,
	axis equal image,
	xmin=1,
	ymin=1,
	ymax=1280,
	xmax=1280,
	xmode=log,
	ymode=log,
	axis x line=bottom,
	axis y line=left,
	xtick={2,4,8,16,32,64,128,256},
	xticklabels={2,4,8,16,,64,,\!\!256},
	extra x ticks = {1,512,724},
	extra x tick labels = {${\le}1$,,\,\,n/a},
	extra x tick style = {grid = major},
	ytick={2,4,8,16,32,64,128,256},
	yticklabels={2,4,8,16,32,64,,},
	extra y ticks = {1,512,724},
	extra y tick labels = {${\le}1$,$\ge$512},
	extra y tick style = {grid = major},
	xlabel={#3},
	xlabel style={font=\scriptsize,yshift=5pt},%{yshift=16pt},
	ylabel={#5},
	ylabel style={font=\scriptsize,yshift=-9pt},%{yshift=-0.4cm},
	yticklabel style={font=\scriptsize},
	xticklabel style={font=\scriptsize},%rotate=290,anchor=west,
	legend pos=north east,
	legend columns=3,
	legend style={nodes={scale=0.75, transform shape},inner sep=1.5pt, xshift=1mm, yshift=7mm},
	set layers,
	mark layer=axis background
	]
	
	\iftoggle{showplots}{\addplot[
	scatter,
	only marks,
	scatter/classes={
		ma={mark=square*,color1,mark size=1.25},
		mdp={mark=triangle*,color2,mark size=1.75},
		pta={mark=o,color3,mark size=1.75}
	},
	scatter src=explicit symbolic
	]%
	table [col sep=tab,x=#2,y=#4,meta=Type] {#1};
	}{\node[anchor=south west, align=center, red] {\huge NOT\\ COMPILED};}
	\ifthenelse{\NOT\equal{#6}{false}}{\legend{MA, MDP,PTA}}{}
	\addplot[no marks] coordinates {(0.01,0.01) (512,512) };
	\addplot[no marks, densely dotted] coordinates {(0.01,0.02) (256,512)};
	\addplot[no marks, densely dotted] coordinates {(0.02,0.01) (512,256)};
	\end{axis}
	\end{tikzpicture}
}

%% file: 1_introduction.tex
\section{Introduction}
\vspace{-0.3cm}
The verification of MDPs is crucial for the design and evaluation of cyber-physical systems with sensor noise, biological and chemical processes, network protocols, and many other complex systems.
MDPs are the standard model for sequential decision making under uncertainty and thus at the heart of reinforcement learning.
Many dependability evaluation and safety assurance approaches rely in some form on the verification of MDPs with respect to temporal logic properties.
Probabilistic model checking~\cite{DBLP:reference/mc/BaierAFK18,DBLP:series/lncs/BaierHK19} provides powerful tools to support this task. %, among others.

The essential MDP model checking queries are for the \emph{worst-case probability that something bad happens} (reachability) and the \emph{expected resource consumption until task completion} (expected rewards).
These are \emph{indefinite (undiscounted) horizon} queries:
They ask about the probability of or the expectation of a random variable up until an event---which forms the horizon---but are themselves unbounded.
Many more complex properties internally reduce to solving either reachability or expected rewards.
For example, if the description of \emph{something bad} is in linear temporal logic (LTL), then a product construction with a suitable automaton reduces the LTL query to reachability~\cite{BK08}.
This paper sets out to determine the practically best algorithms to solve indefinite horizon reachability probabilities and expected rewards; our methodology is an empirical evaluation.
%Our methodology is an empirical evaluation of the algorithmic choices made within the prevalent families of algorithms.

MDP analysis is well studied in many fields and has lead to three main types of algorithms:
\emph{value iteration} (VI), \emph{policy iteration} (PI), and \emph{linear programming} (LP)~\cite{puterman}.
% These families are explained later. 
While indefinite horizon queries are natural in a verification context, they differ from the standard problem of \eg operations research, planning, and reinforcement learning.
In those fields, the primary concern is to \emph{compute a policy} that (often approximately) optimises the \emph{discounted} expected reward over an infinite horizon
where rewards accumulated in the future are weighted by a discount factor $<1$ that exponentially prefers values accumulated earlier.
%Intuitively, such objectives aim to maximise a value accumulated over infinitely long runs.
%To avoid infinite values, rewards accumulated in the future are weighted by a discount factor that exponentially prefers values accumulated earlier during a system run over those accumulated later.
%It is fair to say that such discounting is well-motivated for models of financial profit, where profit today is preferable over profit next year. 

The lack of discounting in verification has vast implications.
The \emph{Bellman operation}, essentially describing a one-step backward update on expected rewards, is a contraction with discounting, but not a contraction without.
This leads to significantly more complex termination criteria for VI-based verification approaches~\cite{HM18}.
Indeed, VI runs in polynomial time for every fixed discount factor~\cite{DBLP:conf/uai/LittmanDK95}, and similar results are known for PI as well as LP solving with the simplex algorithm~\cite{PolPI}.
In contrast, VI~\cite{BK0PS19-complexityVI} and PI~\cite{DBLP:conf/icalp/Fearnley10} are known to have exponential worst-case behaviour in the undiscounted case.

So, \emph{what is the best algorithm for model checking MDPs?}
A polynomial-time algorithm exists using an LP formulation and barrier methods for its solution~\cite{DBLP:books/cu/BV2014}.
LP-based approaches (and their extension to MILPs) are also prominent for more advanced queries such as multi-objective model checking~\cite{DBLP:conf/atva/ForejtKP12}, in counterexample generation~\cite{DBLP:conf/tacas/FunkeJB20}, and for the analysis of parametric Markov chains~\cite{unpublished:tacconvex}.
However, folklore tells us that iterative methods, in particular VI, are better for solving MDPs.
Indeed, variations of VI are the default choice of all model checkers participating in the QComp competition~\cite{DBLP:conf/isola/BuddeHKKPQTZ20}. 
This uniformity may be misleading.
Indeed, for stochastic games and a particular technique, using LP to solve the underlying MDPs may be preferential~\cite[Appendix E.4]{Maxi-ATVA-arxiv}.
For examples in runtime assurance, numerical instability meant that PI was preferred~\cite[Sect. 6]{DBLP:conf/cav/JungesTS20}.
A toy example from~\cite{HM18} is a famous challenge for VI-based methods.
Despite the prominence of LP, the relative ease of encoding MDPs, and the availability of powerful off-the-shelf LP solvers, many model checking tools did (until very recently) not include MDP model checking via LP solvers.
%\cref{tab:intro-comparison} summarises the types of algorithms.

%Did the folklore knowledge mean that probabilistic model checking turned a blind eye to the existence of better solvers. In short, the answer is \emph{not really}.
%\paragraph{Our results.} 
With this paper, we reconsider the PI and LP algorithms to investigate whether probabilistic model checking focused on the wrong family of algorithms.
We report the results of an extensive empirical study with two independent implementations in the model checkers \storm~\cite{HJKQV22} and \mcsta~\cite{HH14}.
%We focus on scalability which is key to enable the analysis of complex systems.
We find that, in terms of performance and scalability, optimistic value iteration~\cite{HK20-ovi} is a solid choice on the standard benchmark collection (which goes beyond competition benchmarks) but can be beat quite considerably on challenging cases.
We also emphasise the question of precision and soundness.
Numerical algorithms, in particular ones that converge \emph{in the limit}, are prone to delivering wrong results.
For VI, the recognition of this problem has led to a series of improvements over the last decade~\cite{BKLPW17,HM18,HK20-ovi,DBLP:conf/cav/KelmendiKKW18,DBLP:conf/cav/PhalakarnTHH20,QK18-svi}.
We show that PI faces a similar problem.
When using floating-point arithmetic, additional issues may arise~\cite{H22,WKHB08}.
%, as was recently highlighted in the context of probabilistic model checking~\cite{H22}.
Our use of various LP solvers exhibits concerning results for a variety of benchmarks. 
We therefore also include results for \emph{exact} computation using rational arithmetic.

\paragraph{Limitations of this study.}
A thorough experimental study of algorithms requires a carefully scoped evaluation.
We work with flat representations of MDPs that fit completely into memory (\ie we ignore the state space exploration process and symbolic methods).
We selected algorithms that are tailored to converge to \emph{the} optimal value.
We also exclude approaches that incrementally build and solve (partial or abstract) MDPs using simulation or model checking results to guide exploration:
they are an orthogonal improvement and would equally profit from faster algorithms to solve the partial MDPs.
Moreover, this study is on algorithms, not on their implementations.
To reduce the impact of potential implementation flaws, we use two independent tools where possible.
Our experiments ran on a single type of machine---we do not study the effect of different hardware.

\paragraph{Contributions.}
This paper contributes a thorough overview on how to model-check indefinite horizon properties on MDPs, making MDP model checking more accessible, but also pushing the state-of-the-art by clarifying open questions. 
Our study is built upon a thorough empirical evaluation using two independent code bases, sources benchmarks from the standard benchmark suite and recent publications, compares 10 LP solvers, and studies the influence of various prominent preprocessing techniques. 
The paper provides new insights and reviews folklore statements:
Particular highlights are a new simple but challenging MDP family that leads to wrong results on all floating-point LP solvers (\Cref{sec:prelim-guarantees}), a negative result regarding the soundness of PI with epsilon-precise policy evaluators (\Cref{sec:zoom-pi}), and an evaluation on numerically challenging benchmarks that shows the limitations of value iteration in a practical setting (\Cref{sec:runtimemonitors}).

% \begin{itemize}
%     \item We optimize LP, using all commonly mentioned tricks and engineering improvements (Sec 5). Moreover, we investigate some ways to improve PI (Sec 6).
%     \item We compare all methods on the QVBS [] (Sec 3), the standard benchmark set for verification tasks. Here, we confirm that VI is best in an approximative setting. However, in an exact setting, PI and LP have their merits and PI wins.
%     \item We investigate challenging benchmarks (Sec 4): some from QVBS, some from Sebastian's paper or Maxi's paper. Again, here VI struggles and PI wins.
% \end{itemize}

%% file: 2_prelims.tex
	\vspace{-0.3cm}
\section{Background}
\vspace{-0.25cm}

We recall MDPs with reachability and reward objectives, describe solution algorithms and their guarantees, and address commonly used optimisations.

\vspace{-0.3cm}
\subsection{Markov Decision Processes}\label{sec:prelim-MDP}
\vspace{-0.25cm}

%A probability distribution over a set $X$ is a mapping $\dist~\colon~X \to [0,1]$ such that for all $x\in X$ we have $0 \geq \dist(x) \geq 1$ and $\sum_{x\in X} \dist(x) = 1$.
%We denote the set of all probability distributions over $X$ as $\dists(X)$.

Let $\dists_X \coloneqq \{\dist \colon X \to [0,1] \mid \sum_{x\in X} \dist(x) = 1\} $ be the set of distributions over $X$.

\begin{definition}
A Markov decision process (MDP) \cite{puterman} is a tuple $\mdp = (\states,\actions,\trans)$ with finite sets of states $\states$ and actions $\actions$, and partially defined transition function $\trans\colon \states \times \actions \to \dists_\states$ such that $\actions(s) \coloneqq \set{a \mid (s,a) \in \mathit{domain}(\delta)} \neq \emptyset$ for all $s \in \states$.

%We assume $\trans$ , i.e.\ not every action is enabled in every state, and that the MDP is non-blocking, i.e.\ in every state at least one action is enabled, formally for all $s \in \states$ there exists $a\in\actions$ such that $\trans(s,a)$ is defined.
\end{definition}
$\actions(s)$ is the set of enabled actions at state $s$. 
$\trans$ maps enabled state-action pairs to distributions over successor states. 
A Markov chain (MC) is an MDP with $|\actions(s)| = 1$ for all $s$.
The \emph{semantics} of an MDP are defined in the usual way, see, \eg \cite[Chapter 10]{BK08}.
%by means of paths and policies.
A (memoryless deterministic) policy---a.k.a. strategy or scheduler---is a function $\policy \colon \states \to \actions$ that, intuitively, given the current state $s$ prescribes what action $a \in \actions(s)$ to play.
Applying a policy $\policy$ to an MDP induces an MC $\mdp^{\policy}$.
A path in this MC is an infinite sequence $\apath = s_1 s_2 \ldots$ with $\trans(s_i, \policy(s_i))(s_{i+1}) > 0$.
$\paths$ denotes the set of all paths and $\prob^\policy_s$ denotes the unique probability measure of $\mdp^\policy$ over infinite paths starting in a state $s$.

%The \textbf{objectives} we consider are reachability and expected rewards. 
A \emph{reachability objective} $\Pobj{\opt}{\targets}$ with set of target states $\targets\subseteq\states$ and $\opt \in \{\max,\min\}$ induces a random variable $X\colon \paths\to[0,1]$ over paths by assigning $1$ to all paths that eventually reach the target and $0$ to all others.
$\Eobj{\opt}{\rew}$ denotes an \emph{expected reward objective}, where $\rew\colon \states \to \Rationals_{\ge 0}$ assigns a reward to each state.
$\rew(\apath) \eqdef \sum_{i=1}^\infty \rew(s_i)$ is the accumulated reward of a path $\apath = s_1 s_2 \dots$.
This yields a random variable $X\colon \paths\to\Rationals \cup \{\infty\}$
that maps paths to their reward.
For a given objective and its random variable $X$, the \emph{value of a state} $s\in\states$ is the expectation of $X$ under the probability measure $\prob^\policy_s$ of the the MC induced by an optimal policy $\policy$, formally
%\footnote{Note that typically $\sup$ and $\inf$ are used instead of $\max$ and $\min$. However, for the objectives we consider, memoryless deterministic policies suffice, see~\cite[Lemma 10.102 and 10.113]{BK08} for reachability and~\cite[Proposition 2]{DBLP:journals/mor/BertsekasT91} for expected reward. Thus, as there are only finitely many of these policies, we can equivalently use $\max$ and $\min$.} 
$\val(s) \eqdef \opt_{\policy\in\policies} \mathbb{E}_s^\policy [X]$.

\vspace{-0.3cm}
\subsection{Solution Algorithms}\label{sec:prelim-algos}
\vspace{-0.25cm}

\emph{Value iteration (VI)}, \eg \cite{CH08}, computes a sequence of value vectors converging to the optimum in the limit.
In all variants of the algorithm, we start with a function $x\colon~\states \to \Rationals$ that assigns to every state an estimate of the value.
% further, we require that $x$ is either a valid under- or over-approximation, i.e.\ for all states $s$ we have either $x(s) \leq \val(s)$ or $x(s) \geq \val(s)$.
The algorithm repeatedly performs an update operation to improve the estimates. 
After some preprocessing, this operation has a unique fixpoint when $x = \val$.
Thus, value iteration converges to the value in the limit.
Variants of VI include interval iteration~\cite{HM18}, sound VI~\cite{QK18-svi} and optimistic VI~\cite{HK20-ovi}.
% We could also mention Gauß-Seidel, but that's more a folklore thing and probably not important.
We do not discuss these in detail, but instead refer to the respective papers.

\emph{Linear Programming (LP)}, \eg \cite[Chapter 10]{BK08}, encodes the transition structure of the MDP and the objective as a linear optimization problem.
For every state, the LP has a variable representing an estimate of its value.
Every state-action pair is encoded as a constraint on these variables, as are the target set or rewards.
The unique optimum of the LP is attained if an only if for every state its corresponding variable is set to the value of the state.
We provide an in-depth discussion of theoretical and practical aspects of LP in \cref{sec:zoom-lp}.

\emph{Policy iteration (PI)}, \eg \cite[Section 4]{DBLP:journals/mor/BertsekasT91}, computes a sequence of policies.
%\footnote{This initial policy needs to be \emph{proper}, i.e.\ it does not induce infinite loops outside of the target or sink states. For a formal definition, see e.g.~\cite[Definition 4]{DBLP:journals/jcss/ChatterjeeAH13} for reachability and~\cite[Section 2]{DBLP:journals/mor/BertsekasT91} for expected reward objectives.}
Starting with an initial policy, we evaluate its induced MC, improve the policy by switching suboptimal choices and repeat the process on the new policy. 
As every policy improves the previous one and there are only finitely many memoryless deterministic policies (a number exponential in the number of states), eventually we obtain an optimal policy.
We further discuss PI in \cref{sec:zoom-pi}.

\vspace{-0.3cm}
\subsection{Guarantees}\label{sec:prelim-guarantees}
\vspace{-0.25cm}

Given the stakes in many model checking applications, we require guarantees about the relation between an algorithm's result $\bar v$ and the true probability or expected reward $v$.
First, implementations are subject to floating-point errors and imprecision unless they use exact (rational) arithmetic or safe rounding~\cite{H22}.
This can result in arbitrary differences between result and true value.
Second are the inherent properties of the algorithm:
VI is an approximating algorithm that converges to the true value only in the limit.
In theory, it is possible to obtain the exact result by rounding after an exponential number of iterations~\cite{CH08}; in practice, this results in excessive runtime.
Instead, for years, implementations used a naive stopping criterion that could return arbitrarily wrong results~\cite{HM14}.
This problem's discovery~\cite{HM14} sparked the development of various sound variants of VI~\cite{BKLPW17,HM18,HK20-ovi,DBLP:conf/cav/KelmendiKKW18,DBLP:conf/cav/PhalakarnTHH20,QK18-svi}, including interval iteration, sound value iteration, and optimistic value iteration.
A sound VI algorithm guarantees $\varepsilon$-precise results, \ie $|v - \bar v| \leq \epsilon$ or $|v - \bar v| \leq v \cdot \epsilon$.
For LP and PI, the question of guarantees has not yet been thoroughly investigated.
Theoretically, both are exact, but implementations are often not.
We discuss the problems in detail in \cref{sec:zoom-lp,sec:zoom-pi}.

\begin{figure}[t]
\begin{minipage}[b]{0.715\linewidth}
\centering
%\tikzset{->,>=stealth'}
\tikzstyle{every node}=[font=\scriptsize]
\begin{tikzpicture}[on grid, auto]
    \node[state] (s0) {$0$};
    \coordinate[left=0.3 of s0.west] (start);
    \node[] (me) [above left=1.1 and 0.5 of s0] {\small$M_n$:};
    \node[dot] (n0) [right=0.75 of s0] {};
    \node[state] (sp1) [above right=0.75 and 1.0 of n0] {$1$};
    \node[state] (sm1) [below right=0.75 and 1.0 of n0] {$\sm1$};
    \node[dot] (nsp1m) [above right=0.4 and 0.75 of sp1] {};
    \node[dot] (nsp1j) [below right=0.2 and 0.85 of sp1] {};
    \node[dot] (nsm1j) [above right=0.2 and 0.85 of sm1] {};
    \node[dot] (nsm1m) [below right=0.4 and 0.75 of sm1] {};
    \node[state] (sp2) [above right=0.4 and 0.75 of nsp1m] {$2$};
    \node[state] (sm2) [below right=0.4 and 0.75 of nsm1m] {$\sm2$};
    \node[] (nsp2m) [above right=0.4 and 0.75 of sp2] {};
    \node[] (nsp2j) [below right=0.2 and 0.85 of sp2] {};
    \node[] (nsm2j) [above right=0.2 and 0.85 of sm2] {};
    \node[] (nsm2m) [below right=0.4 and 0.75 of sm2] {};
    \node[] (spdots) [above right=0.1 and 1.67 of sp2] {$\cdots$};
    \node[] (smdots) [below right=0.1 and 1.67 of sm2] {$\cdots$};
    \node[state,double] (spn) [right=4.75 of sp1] {$n$};
    \node[state] (smn) [right=4.75 of sm1] {$\sm{n}$};
    \node[] (nspxm) [above left=1 and 0.75 of spn] {};
    \node[] (nsmxm) [below left=1 and 0.75 of smn] {};
    ;
    \path[-]
      (s0) edge[] node[pos=0.5,inner sep=3pt] {$\tau$} (n0)
      (sp1) edge[pos=0.67] node[inner sep=2pt] {\texttt{m}} (nsp1m)
      (sp1) edge[pos=0.67,swap] node[inner sep=2pt] {\texttt{j}} (nsp1j)
      (sm1) edge[pos=0.67] node[inner sep=2pt] {\texttt{j}} (nsm1j)
      (sm1) edge[pos=0.67,swap] node[inner sep=2pt] {\texttt{m}} (nsm1m)
    ;
    \path[-,dashed]
      (sp2) edge[pos=0.67] node[inner sep=2pt] {\texttt{m}} (nsp2m)
      (sp2) edge[pos=0.67,swap] node[inner sep=2pt] {\texttt{j}} (nsp2j)
      (sm2) edge[pos=0.67] node[inner sep=2pt] {\texttt{j}} (nsm2j)
      (sm2) edge[pos=0.67,swap] node[inner sep=2pt] {\texttt{m}} (nsm2m)
    ;
    \path[->]
      (start) edge node {} (s0)
      (n0) edge [] node[inner sep=1pt] {$\frac{1}{2}$} (sp1)
      (n0) edge [] node[swap,inner sep=1pt] {$\frac{1}{2}$} (sm1)
      (nsp1m) edge [] node[swap,pos=0.33,inner sep=1.0pt] {$\frac{1}{2}$} (sp2)
      (nsp1m) edge [out=90,in=70] node[pos=0.25,right] {~~$\frac{1}{2}$} (s0)
      (nsm1m) edge [] node[pos=0.33,inner sep=1.0pt] {$\frac{1}{2}$} (sm2)
      (nsm1m) edge [out=-90,in=-70] node[pos=0.25,right] {~~$\frac{1}{2}$} (s0)
      (nsp1j) edge [out=10,in=160] node[pos=0.7,below,inner sep=2pt] {$\frac{1}{2}$} (spn)
      (nsp1j) edge [out=-10,in=120] node[pos=0.1,below,inner sep=2pt] {$\frac{1}{2}$} (smn)
      (nsm1j) edge [out=10,in=-120] node[pos=0.1,above,inner sep=2pt] {$\frac{1}{2}$} (spn)
      (nsm1j) edge [out=-10,in=-160] node[pos=0.7,above,inner sep=2pt] {$\frac{1}{2}$} (smn)
      (spn) edge [loop,out=30,in=-30,looseness=5] node[right,pos=0.5,inner sep=2pt] {$1$} (spn)
      (smn) edge [loop,out=-30,in=30,looseness=5] node[right,pos=0.5,inner sep=2pt] {$1$} (smn)
    ;
    \path[->,dashed]
      (nspxm) edge [bend left] node[pos=0.5,inner sep=1.5pt] {$\frac{1}{2}~\cdots$} (spn)
      (nsmxm) edge [bend right] node[swap,pos=0.5,inner sep=1.5pt] {$\frac{1}{2}~\cdots$} (smn)
    ;
\end{tikzpicture}
\caption{A hard MDP for all algorithms}
\label{fig:hmmdp}
\end{minipage}%
\begin{minipage}[b]{0.285\linewidth}
\centering
\captionof{table}{Correct results}
\label{tab:hmmdp}
\setlength{\tabcolsep}{6pt}
\begin{tabular}[b]{llc}\toprule
alg. & solver & $n\,{\leq}$\\\midrule
PI & -- & $20$\\\midrule
LP & \solver{COPT} & $18$\\
   & \solver{CPLEX} & $18$\\
   & \solver{Glop} & $25$\\
   & \solver{GLPK} & $24$\\
   & \solver{Gurobi} & $18$\\
   & \solver{HiGHS} & $22$\\
   & \solver{lp\_solve} & $28$\\
   & \solver{Mosek} & $22$\\
   & \solver{SoPlex} & $34$
\\\bottomrule
\end{tabular}
\end{minipage}
\end{figure}

The handcrafted MC of \cite[Figure 2]{HM14} highlights the lack of guarantees of VI:
standard implementations return vastly incorrect results.
We extended it with action choices to obtain the MDP $M_n$ shown in \Cref{fig:hmmdp} for $n \in \mathbb{N}$, $n \geq 2$.
It has $2n+1$ states; we compute $\Pobj{\min}{\set{n}}$ and $\Pobj{\max}{\set{n}}$.
The policy that chooses action \texttt{m} wherever possible induces the MC of \cite[Figure 2]{HM14} with $\tuple{\Pobj{\min}{\set{n}}, \Pobj{\max}{\set{n}}} = \tuple{\frac{1}{2}, \frac{1}{2}}$.
In every state $s$ with $0 < s < n$, we added the choice of action \texttt{j} that jumps to $n$ and $\sm{n}$.
With that, the (optimal) values over all policies are $\tuple{\frac{1}{3}, \frac{2}{3}}$.
In VI, starting from value $0$ for all states except $n$, initially taking \texttt{j} everywhere looks like the best policy for \Pmax.
As updated values slowly propagate, at some point, state-by-state, \texttt{m} becomes the optimal choice.
We thus layered a ``deceptive'' decision problem on top of the slow convergence of the original MC.
Consequently, for $n = 20$, VI with \storm and \mcsta delivers the incorrect results $\tuple{0.247, 0.500}$.
For \storm's PI and various LP solvers, we show in \Cref{tab:hmmdp} the largest $n$ for which they return a $\pm\,0.01$-correct result.
For larger $n$, PI and all LP solvers claim $\approx \tuple{\frac{1}{2}, \frac{1}{2}}$ as the correct solution except for \solver{Glop} and \solver{GLPK}, which return $\approx \tuple{\frac{1}{3}, \frac{1}{2}}$ until giving up for the minimum at $n = 29$ and $52$, respectively.
Sound VI algorithms and \storm's exact-arithmetic engine produce ($\epsilon$-)correct results, though the former at excessive runtime for larger $n$.
We used default settings for all tools and solvers.

\vspace{-0.3cm}
\subsection{Optimizations}\label{sec:prelim-opts}
\vspace{-0.25cm}

VI, LP, and PI can all benefit from the following optimizations.

%\begin{itemize}
    %\item \
    \paragraph{Graph-theoretic algorithms} can be used for qualitative analysis of the MDP, i.e.\ finding states with value 0 or (only for reachability objectives) 1. These qualitative approaches are typically a lot faster than the numerical computations for quantitative analysis. Thus, we always apply them first and only run the numerical algorithms on the remaining states with non-trivial value.
    % \textcolor{blue}{For reward, there are no value 1 states. Our definition of reward prevents value $\infty$ states; thus, we do not have to say sth like "min and max value", but can leave this and just use the fact that for reward, there are no value 1, but we find the value 0 ones.}
   % \item The \textbf
   
    \paragraph{Topological methods}, \eg \cite{TVI}, does not consider the whole MDP at once. Instead, we first compute a topological ordering of the strongly connected components (SCCs)\footnote{A set $\states' \subseteq \states$ is a connected component if for all $s,s' \in \states'$ s can be reached from s'. We call $\states'$ strongly connected component if there is no superset of $\states'$ that is a connected component.} and then analyze each SCC individually. This can improve the runtime, as we decompose the problem into smaller subproblems.
    The subproblems can be solved with any of the solution methods.
    Note that when considering acyclic MDPs, the topological approach does not need to call the solution methods, as the resulting values can immediately be backpropagated.
    
\paragraph{Collapsing of maximal end components (MECs)}%
%\footnote{A set of states $\states' \subseteq \states$ is an end component, if there exists a policy $\policy$ such that in the MC induced by $\policy$, from all states $s\in\states'$ the probability to reach every other state $t_1\in\states'$ is positive and the probability to reach some state $t_2\in\states\setminus\states'$ outside of the end component is zero. We call $\states'$ maximal end component if there is no superset of $\states'$ that is an end component.}
, \eg,~\cite{BCC+14,HM18}, transforms the MDP into one with equivalent values but simpler structure. 
    After collapsing MECs, the MDP is contracting, \ie we almost surely reach a target state or a state with value zero. 
    VI algorithms rely on this property for convergence~\cite{HM18,QK18-svi,HK20-ovi}. For PI and LP, simplifying the graph structure before applying the solution method can speed up the computation.
%Our experiments in \Cref{sec:qcomp} show that the overhead for collapsing usually is equal to the time saved.
    % \item \textbf{Simplifying the property} makes the MDP solving easier. This includes (i) only requiring to compute the value in a given initial state, not for all states and (ii) not computing the precise value, but only deciding whether the value is greater or smaller than some given threshold. 
    % We comment on this optimization in \cref{sec:zoom-lp}, but do not provide a thorough analysis and comparison for all algorithms, because the impact greatly depends on the chosen initial state, threshold and $\opt$, the direction of optimization.
    % \todo{Update this once we know what we say in 3}
%    \todo{I commented out stuff on simplifying the property}

    \paragraph{Warm starts}, \eg \cite{Gir14,KRSW22}, may adequately initialize an algorithm, \ie, we may provide it with some prior knowledge so that the computation has a good starting point.
We implement warm starts by first running VI for a limited number of iterations and using the resulting estimate to guess bounds on the variables in an LP or a good initial policy for PI. 
	See Sections~\ref{sec:zoom-lp} and~\ref{sec:zoom-pi} for more details.
    
    % \todo{Potential sentence on VI in comment here}
    %VI can also benefit from prior knowledge on the estimate vector, but it is unclear where this knowledge comes from in general; still, in a sense optimistic VI first performs VI to guess an upper estimate vector that is closer to optimal, so from a certain viewpoint it is interval iteration with a warm start

%% file: 3_zoomLP.tex
\vspace{-0.3cm}
\section{Practically solving MDPs using Linear Programs}\label{sec:zoom-lp}
\vspace{-0.25cm}

This section considers the LP-based approach to solving the optimal policy problem in MDPs.
To the best of our knowledge, this is the only polynomial time approach.
We discuss various configurations. These configuration are a combination of the LP formulation, the choice of software, and their parameterization.  

\vspace{-0.3cm}
\subsection{How to encode MDPs as LPs?}
\vspace{-0.25cm}

For objective $\Pobj{\max}{\targets}$ we formulate the following LP over variables $x_s$, $s \in \states \setminus \targets$: 
\begin{align*}
     \text{minimize}\quad&  \sum_{s \in \states} x_s \quad\text{s.t. }  \lowerbound(s) \leq x_s \leq \upperbound(s) \quad\text{and} \\
     \quad&  x_s \geq \sum_{s' \in \states \setminus \targets}\trans(s,a)(s') \cdot  x_{s'} + \sum_{t \in \targets} \trans(s,a)(t) \quad \text{ for all } s \in \states \setminus \targets, a \in \actions \\
     %\quad&  \lowerbound(s) \leq x_s \leq \upperbound(s) 
\end{align*}
We assume bounds $\lowerbound(s) = 0$ and $\upperbound(s) = 1$ for $s \in \states \setminus \targets$. %and $\lowerbound(t) = \upperbound(t) = 1$ for $t \in \targets$.
The unique solution $\eta \colon \set{x_s \mid s \in \states \setminus \targets} \to [0,1]$ to this LP coincides with the desired objective values $\eta(x_s) = V(s)$.
Objectives $\Pobj{\min}{\targets}$ and $\Eobj{\opt}{\rew}$ have similar encodings: minimising policies require maximisation in the LP and flipping the constraint relation. Rewards can be added as an additive factor on the right-hand side. 
For practical purposes, the LP  formulation can be tweaked. 

%This immediately yields a polynomial time algorithm for solving MDPs. 

\paragraph{The choice of bounds.}
Any bounds that respect the unique solution will not change the answer. That is, any $\lowerbound$ and $\upperbound$ with $0 \leq \lowerbound(s) \leq V(s) \leq \upperbound(s)$ yield a sound encoding.
While these additional bounds are superfluous, they may significantly prune the search space. We investigate trivial bounds, e.g., knowing that all probabilities are in $[0,1]$, bounds from a structural analysis as discussed by~\cite{BKLPW17}, and bounds induced by a warm start of the solver.
For the latter, if we have obtained values $V' \leq V$, e.g., induced by a suboptimal policy, then $V'(s)$ is a lower bound on the value $x_s$, which is particularly relevant as the LP minimizes.

\paragraph{Equality for unique actions.}
Markov chains, i.e., MDPs where $|\actions| = 1$, 
can be solved using linear equation systems. The LP encoding uses one-sided inequalities and the objective function to incorporate nondeterministic choices. We investigate adding constraints for all states with a unique action.
%, i.e., to all states $S_\text{det} = \{ s \mid |A(s)| = 1\}$.
\[ x_s \leq \sum_{s' \in S \setminus T}\trans(s,a)(s') \cdot  x_{s'} + \sum_{t \in T} \trans(s,a)(t) \quad \text{ for all } s \in S \setminus T \text{ with } \actions(s) = \{a\}\]
These additional constraints may trigger different optimizations in a solver, e.g., some solvers use Gaussian elimination for variable elimination.

\paragraph{A simpler objective.}
The standard objective assures the solution $\eta$ is optimal for \emph{every} state, whereas most invocations require only optimality in some specific states -- typically the initial state $s_0$ or the entry states of a strongly connected component.
In that case, the objective may be simplified to optimize only the value for those states. 
This potentially allows for multiple optimal solutions: in terms of the MDP, it is no longer necessary to optimize the value for states that are not reached under the optimal policy.
%It is a priori unclear whether this is helpful in practice.
% On one hand, this reduction means that more feasible solutions of the LP also solve the LP\footnote{}. It thus makes it easier to solve the LP. 

\paragraph{Encoding the dual formulation.}
Encoding a dual formulation to the LP is interesting for mixed-integer extensions to the LP,  relevant for computing, e.g., policies in POMDPs~\cite{kumar2015history}, or when computing minimal counterexamples~\cite{DBLP:journals/corr/abs-1305-5055}. For LPs, due to the strong duality, the internal representation in the solvers we investigated is (almost) equivalent and all solvers support both solving the primal and the dual representation. We therefore do not further consider constructing them. 

\vspace{-0.3cm}
\subsection{How to solve LPs with existing solvers?}
\vspace{-0.25cm}

We rely on the performance of state-of-the-art LP solvers. Many solvers have been developed and are still actively advanced, see~\cite{anand2017comparative} for a recent comparison on general benchmarks.
We list the LP solvers that we consider for this work in  \cref{tab:lpsolvers}.
The columns summarize for each solver the type of license, whether it uses exact or floating point arithmetic, whether it supports multithreading, and what type of algorithms it implements. We also list whether the solver is available from the two model checkers used in this study\footnote{Support for \solver{Gurobi}, \solver{GLPK}, and \solver{Z3} was already available in \storm. Support for \solver{Glop} was already available in \mcsta. All other solver interfaces have been added.}.

\paragraph{Methods.}
We briefly explain the available methods and refer to~\cite{DBLP:books/cu/BV2014} for a thorough treatment.
Broadly speaking, the LP solvers use one out of two families of methods. \emph{Simplex}-based methods rely on highly efficient pivot operations to consider vertices of the simplex of feasible solutions. 
Simplex can be executed either in the \emph{primal} or \emph{dual} fashion, which changes the direction of progress made by the algorithm. Our LP formulation has more constraints than variables, which generally means that the dual version is preferable. \emph{Interior methods}, often the subclass of \emph{barrier methods}, do not need to follow the set of vertices. These methods may achieve polynomial time worst-case behaviour. It is generally claimed that simplex has superior average-case performance but is highly sensitive to perturbations, while interior-point methods have a more robust performance.

\paragraph{Warm starts.}
LP-based model checking can be done using two types of warm starts. Either by providing a (feasible) basis point as done in \cite{Gir14} or by presenting bounds. 
The former, however, comes with various remarks and limitations, such as the requirement to disable preprocessing. We therefore used warm starts only by using bounds as discussed above. 

\paragraph{Multithreading.}
We generally see two types of parallelisation in LP solvers. Some solver support a \emph{portfolio} approach that run different approaches and finishes with the first one that yields a result. Other solvers parallelise the interior-point and/or simplex methods themselves.

\newcounter{mpFootnoteValueSaver}
\setcounter{mpFootnoteValueSaver}{\value{footnote}}

\begin{table}[t]
	\centering
	\caption{Available LP solvers {\footnotesize(``intr'' = interior point)}}
    \vspace{6pt}
    \setlength{\tabcolsep}{4pt}
	\begin{tabular}{lclccccc}
		\toprule
		solver & version & license & exact/fp & parallel & algorithms & \mcsta & \storm   \\\midrule
		\solver{CPLEX}\footnotemark & 22.10 & academic & fp & yes & intr\,+\,simplex  & yes & no \\
		\solver{COPT}\footnotemark & 5.0.5 & academic & fp & yes & intr\,+\,simplex & yes & no  \\
		\solver{Gurobi}~\cite{gurobi} & 9.5  & academic & fp & yes & intr\,+\,simplex & yes & yes \\
		\solver{GLPK}\footnotemark & 4.65 & GPL &  fp & no & intr\,+\,simplex & no & yes  \\
		\solver{Glop}\footnotemark & 9.4.1874 & Apache & fp & no & simplex only & yes & no \\
		\solver{HiGHS}\footnotemark & 1.2.2 & MIT & fp & yes & intr\,+\,simplex  & yes & no   \\
		%highs explicitly asks to cite this, but the publication is not clearly related to highs, which is why I prefer the website. IMO, we can also drop the citation.
		\solver{lp\_solve}\footnotemark & 5.5.2.11 & LGPL & fp & no & simplex only & yes & no \\ 
		\solver{Mosek}\footnotemark & 10.0 & academic &  fp & yes & intr\,+\,simplex & yes & no \\
		\solver{SoPlex}~\cite{soplex} & 6.0.1  & academic & both & no & simplex only & no & yes \\
		\solver{Z3}~\cite{z3} & 4.8.13 & MIT & exact & no & simplex only & no & yes \\
		\bottomrule
	\end{tabular}
	\label{tab:lpsolvers}
\end{table}

%Necessary, since footnotes cannot be displayed inside floats.
% https://www.overleaf.com/learn/latex/Footnotes#The_.5Cfootnotemark_and_.5Cfootnotetext_commands

\stepcounter{mpFootnoteValueSaver}
\footnotetext[\value{mpFootnoteValueSaver}]{\url{https://www.ibm.com/analytics/cplex-optimizer}}
\stepcounter{mpFootnoteValueSaver}
\footnotetext[\value{mpFootnoteValueSaver}]{\url{https://www.shanshu.ai/copt}}
\stepcounter{mpFootnoteValueSaver}
\footnotetext[\value{mpFootnoteValueSaver}]{\url{https://www.gnu.org/software/glpk/}}
\stepcounter{mpFootnoteValueSaver}
\footnotetext[\value{mpFootnoteValueSaver}]{\url{https://developers.google.com/optimization/lp}}
\stepcounter{mpFootnoteValueSaver}
\footnotetext[\value{mpFootnoteValueSaver}]{\url{https://www.maths.ed.ac.uk/hall/HiGHS/}}
\stepcounter{mpFootnoteValueSaver}
\footnotetext[\value{mpFootnoteValueSaver}]{\url{https://lpsolve.sourceforge.net/5.5/}}
\stepcounter{mpFootnoteValueSaver}
\footnotetext[\value{mpFootnoteValueSaver}]{\url{https://www.mosek.com/}}

% \subsubsection{Warm starts}
% \color{blue}
% Finally, the single paper investigating LP for MDPs problems we are aware of~\cite{Gir14} already improved the performance of LP significantly. 
% Side note: if we try the \cite{Gir14} idea (basically VI2LP), the LP solvers seem to just discard the additional info. (But i think that one must `just` disable preprocessing. I find it somewhat tricky to make claims about this without backing it up with very hard numbers.

% - Make sure to avoid redundancy between here, the para about warm start in 2.4 and the para about bounds in 3.1. 
% - Do we want to discuss that Giro is kinda using a different warm start, since he picks a point on simplex (a "basis" in his words, I hope I understood that correctly), while we only prune the search space? And is it maybe kinda the same?
% - Should there be a para about warm starts both in 3.1 and 3.2? If so, we gotta make clear what the difference is. I think I vote for just having it in 3.1, since the discussion is more abstract and theoretical.

%\paragraph{Some details on \solver{Gurobi}.}
%\solver{Gurobi} is the fastest numerical LP solver in our experiments. It provides an automatic selection of methods, but one can enforce setting a specific method, being either primal or dual simplex or a barrier method.  Multithreading  follows a portfolio approach. The run times are rather nondeterministic. Some measures can be taken to reduce variability at the cost of performance.
%\todo{Sebastian: drop the gurobi paragraph? Maxi: Agree.}

\paragraph{Guarantees for numerical LP solvers.}
All LP solvers allow tweaking of various parameters, including \emph{tolerances} to manage whether a point  is considered feasible or optimal, respectively. However, the experiments in Tab.~\ref{tab:hmmdp} already indicate that these guarantees are \emph{not} absolute. A limited experiment indicated that reducing these tolerances towards zero did remove some incorrect results, but not all.

\paragraph{Exact solving.}
\solver{Soplex} supports exact computations, with a Boost library wrapping GMP rational numbers\footnote{\url{https://gmplib.org/}}, while exploiting floating point arithmetic~\cite{DBLP:conf/issac/GleixnerSW12}. While this is beneficial for performance in most settings, it seems to raise errors for the numerical challenging models. 
\solver{Z3} supports only exact arithmetic (also wrapping GMP numbers with their own interface). We observe that the price of converting large rational numbers may be substantial.
Furthermore, SMT-solvers like \solver{Z3} use a simplex variation~\cite{DBLP:conf/cav/DutertreM06} tailored towards finding feasible points and in an incremental fashion, optimized for problems with a nontrivial Boolean structure. In contrast, our LP formulation is easily feasible and are a pure conjunction. 

%\begin{remark}\todo{Can be removed?}
%We notice that some advanced model checking queries require solving many similar MDPs. Some solvers support some incrementality in their data structures. E.g,~\cite{unpublished:tacconvex} uses incremental calls to \solver{Gurobi} to avoid rebuilding constraints, this seems to mostly reduce the time to construct the data structures but not to reduce the time for solving. In a (limited) set of examples, we found only little effects in keeping sets of constraints.
%\end{remark}

% \begin{enumerate}
%     \item Primal vs. dual (no difference for us, but \cite{Gir14} had an order of magnitude).
%     \item Different solvers (Soplex, Gurobi, z3, GLPK in Storm. CPLEX is too badly documented. Gurobi, glob, lpsolve and chinese thing [Mindopt: \url{https://solver.damo.alibaba.com/htmlpages/page#/en}] to be in Modest)
%     \item Bounds on the variables ([0,1] for probabilities, Baier et al. for rewards, Available in storm)
%     \item Interface / Representation [Memory footprint (avoid having two representations)] (we have not discussed this yet. Comparing the memory of all the algorithms is interesting, too.)
%     \item Multithreading: LP solver or SCC-decomposition (we have not discussed this idea after initially writing it down). Gurobi can do multithreading. 
% \end{enumerate}

% Can we show plots to say that these things don't make a difference already here? 

%% file: 4_zoomPI.tex
\vspace{-0.3cm}
\section{Sound Policy Iteration}\label{sec:zoom-pi}
\vspace{-0.25cm}

Starting with an initial policy, PI-based algorithms iteratively improve the policy based on the values obtained for the induced MC.
The algorithm for solving the induced MC crucially affects performance and accuracy of the overall approach.
This section addresses the solvers available in \storm, possible precision issues and how to utilize a warm start, while \cref{sec:qcomp} discusses PI performance\footnote{\cite{KRSW22} addresses performance in the context of PI for stochastic games.}.

\paragraph{Markov Chain Solvers.}
%\label{sec:PI:MCSolvers}
%
To solve the induced MC, \storm can employ all linear equation solvers listed in~\cite{HJKQV22} and all variants of VI implemented in \storm.
In our experiments, we consider 
(i)~the generalised minimal residual method (GMRES)~\cite{Saad1986GMRESAG} implemented in \tool{GMM++}\footnote{\url{https://getfem.org/gmm.html}},
(ii)~VI~\cite{CH08} with a standard (relative) termination criterion,
(iii)~optimistic VI (OVI)~\cite{HK20-ovi}, and
(iv)~the sparse LU decomposition implemented in \tool{Eigen}\footnote{\url{https://eigen.tuxfamily.org/index.php}} using either floating point or exact arithmetic (LU$^\mathrm{X}$);
LU and LU$^\mathrm{X}$ provide exact results (modulo floating point errors in LU) OVI yields $\varepsilon$-precise results. VI and GMRES do not provide any guarantees.

% Recall that PI works by fixing a policy, solving the induced system and then greedily updating the policy.
% As already observed in~\cite[Section 4.2]{KRSW22}, the choice of algorithm for solving the induced system is relevant. In that paper, the authors considered stochastic games and found that the different algorithms for solving MDPs affect the runtime of PI.
% We provide such a runtime comparison for different Markov chain solvers when using PI on MDPs in Section~\ref{sec:qcomp}.
% Here, we compare the different Markov chain solutions approaches with respect to their guarantees.

\paragraph{Correctness of PI.}
The accuracy of PI is affected by the MC solver:
Firstly, PI cannot be more  precise as its underlying solver: the result of PI has the same precision as the result obtained for the final MC.
Secondly, inaccuracies by the solver can hide policy improvements which may lead to premature convergence with a sub-optimal policy.
The example below shows that PI can return arbitrarily wrong results---\emph{even if the intermediate results are $\varepsilon$-precise}.

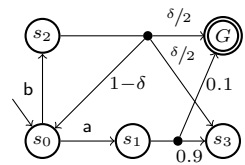
\begin{wrapfigure}[7]{r}{0.29\linewidth}
\vspace{-2.8em} % needed when the next paragraph is not the first on a page
\centering
\input{4_PI-example}%
\caption{Example MDP}
\label{fig:pi:fails}
\end{wrapfigure}
%\begin{example}\label{ex:pi:fails}
Consider the MDP in \cref{fig:pi:fails} with objective $\Pobj{\max}{\set{G}}$.
There is only one nondeterministic choice, namely in state $s_0$.
The optimal policy is to pick $\mathsf{b}$, obtaining a value of $0.5$.
Picking $\mathsf{a}$ only yields $0.1$.
However, when starting from the initial policy $\policy(s_0)=\mathsf{a}$, 
an $\varepsilon$-precise MC solver may return $0.1 + \varepsilon$ for both $s_0$ and $s_1$ and $\sfrac \delta 2 + (1-\delta)\cdot 0.1$ for $s_2$. 
This solution is indeed $\varepsilon$-precise.
However, when evaluating which action to pick in $s_0$, we can choose $\delta$ such that $\mathsf{a}$ seems to obtain a higher value. Concretely, we require $\sfrac \delta 2 + (1-\delta)\cdot 0.1 < 0.1 + \varepsilon$. 
For every $\varepsilon > 0$, this can be achieved by setting $\delta < 2.5 \cdot \varepsilon$.
In this case, PI would terminate with the final policy inducing a severly suboptimal value. 

%\end{example}

% Firstly, the correctness of the intermediate iterations can affect the overall soundness of the algorithm and make the result arbitrarily wrong. We discuss this in more detail in the following paragraphs.
% Secondly, assuming the intermediate iterations are correct, the guarantees of the Markov chain solver in the final iteration carry over to the result of PI on the MDP. For example, if the final Markov chain is solved with $\varepsilon$-precision, the result of PI also is only $\varepsilon$-precise.

% Now we explain four different ways in which the guarantees of the intermediate iterations impact the overall soundness of the algorithm.
% Naturally, i
% If every Markov chain is solved precisely, the algorithm is correct
% ; and if the Markov chain is solved by a heuristic without guarantees, the the algorithm is unsound.

If every Markov chain is solved precisely, PI is correct. Indeed, it suffices to be certain that one action is better than all others. 
This is the essence of modified policy iteration as described in~\cite[Chapters 6.5 and 7.2.6]{puterman}. 
Similarly,~\cite[Section 4.2]{KRSW22} suggests to use interval iteration when solving the system induced by the current policy and stopping when the under-approximation of one action is higher than the over-approximation of all other actions. %(or the other way round in the case of minimization).

\paragraph{Warm starts.}
PI profits from providing a \emph{good} initial policy.
If the initial policy is already optimal, PI terminates after a single iteration.
We can inform our choice of the initial policy by providing estimates for all states as computed by VI. 
For every state, we choose the action that is optimal according to the estimate.
This is a good way to leverage VI's ability to quickly deliver good estimates~\cite{HK20-ovi}, while at the same time providing the exactness guarantees of PI.

%% file: 4_PI-example.tex
\tikzstyle{every node}=[font=\scriptsize]
\begin{tikzpicture}[scale=0.8]

\node[state] (s0) at (0,-1) {$s_0$};
\draw[->] ($(s0) + (-0.5,0.7)$) to  (s0);
\node[state] (s1) at (1.5,-1) {$s_1$};
\node[state] (s2) at (0,0.75) {$s_2$};
\node[state,double] (goal) at (3,0.75) {$G$};
\node[state] (sink) at (3,-1) {$s_3$};

\node[dot] (s1mid) at (2.25,-1) {};
\node[dot] (s2mid) at (1.75,0.75) {};

\draw[->] (s0) to node [anchor=south] {$\mathsf{a}$} (s1);
\draw[->] (s0) to node [anchor=east] {$\mathsf{b}$} (s2);

\draw[-] (s1) to (s1mid);
\draw[->] (s1mid) to node [pos=0.7,anchor=north west,inner sep=0.1pt] {$0.1$} (goal);
\draw[->] (s1mid) to node [pos=0.3,anchor=north] {$0.9$} (sink);

\draw[-] (s2) to (s2mid);
\draw[->] (s2mid) to node [pos=0.5,anchor=south] {$\sfrac \delta 2$} (goal);
\draw[->] (s2mid) to node [pos=0.3,anchor=south west,inner sep=0.1pt] {$\sfrac \delta 2$} (sink);
\draw[->] (s2mid) to node [pos=0.5,anchor=west] {$1{-}\delta$} (s0);

\end{tikzpicture}%

%% file: 5_qcomp.tex
\section{Experimental Evaluation}\label{sec:qcomp}

To understand the practical performance of the different algorithms, we performed an extensive experimental evaluation. We used three sets of benchmarks: all applicable benchmark instances\footnote{A \emph{benchmark instance} is a combination of model, parameter valuation, and objective.} from the Quantitative Verification Benchmark Set (QVBS)~\cite{HKPQR19} (the \textit{qvbs} set), a subset of hard QVBS instances (the \textit{hard} set), and numerically challenging models from a runtime monitoring application~\cite{DBLP:conf/cav/JungesTS20} (the \textit{premise} set, named for the corresponding prototype). We consider two probabilistic model checkers, \storm~\cite{HJKQV22} and the \tool{Modest Toolset}'s~\cite{HH14} \mcsta.
We used Intel Xeon Platinum 8160 systems running 64-bit CentOS Linux~7.9, allocating 4 CPU cores and 32\,GB RAM per experiment unless noted otherwise.

We plot algorithm runtimes in seconds in \emph{quantile plots} as on the left and \emph{scatter plots} as on the right of \Cref{fig:LPSolvers}.
The former compares multiple tools or configurations; for each, we sort the instances by runtime and plot the corresponding monotonically increasing line.
Here, a point $\tuple{x, y}$ on the $a$-line means that the $x$-th fastest instance solved by $a$ took $y$ seconds.
The latter compares two tools or configurations.
Each point $\tuple{x, y}$ is for one benchmark instance: the x-axis tool took $x$ while the y-axis tool took $y$ seconds to solve it. The shape of points indicates the model type; the mapping from shapes to types is the same for all scatter plots and is only given explicitly in the first one in \Cref{fig:LPSolvers}.
Additional plots to support the claims in this section are provided in the appendix.

The depicted runtimes are for the respective algorithm and all necessary and/or stated preprocessing, but do not include the time for constructing the MDP state spaces (which is independent of the algorithms).
\mcsta reports all time measurements rounded to multiples of $0.1$\,s.
We summarise timeouts, out-of-memory, errors, and incorrect results as ``n/a''.
Our timeout is 30 minutes for the algorithm and 45 minutes for total runtime including MDP construction.
We consider a result $\bar v$ incorrect if $|v - \bar v| > v \cdot 10^{-3}$ (\ie relative error $10^{-3}$) whenever a reference result $v$ is available.
We however do not flag a result as incorrect if $v$ and $\bar{v}$ are both below $10^{-8}$ (relevant for the \textit{premise} set).
Nevertheless, we configure the (unsound) convergence threshold for VI as $10^{-6}$ relative; among the sound VI algorithms, we include OVI, with a (sound) stopping criterion of relative $10^{-6}$ error.
To only achieve the $10^{-3}$ precision we actually test, OVI could thus be even faster than it appears in our plots.
We make this difference to account for the fact that many algorithms, including the LP solvers, do not have a sound error criterion.
We mark exact algorithms/solvers that use rational arithmetic with a superscript~$^\mathrm{X}$.
The other configurations use floating points (fp).

\subsection{The QVBS Benchmarks}\label{sec:exp-qvbs}

The \textit{qvbs} set comprises all QVBS benchmark instances with an MDP, Markov automaton (MA), or probabilistic timed automaton (PTA) model\footnote{MA and PTA are converted to MDP via embedding and digital clocks~\cite{KNPS06}.} and a reachability or expected reward/time objective that is quantitative, \ie not a query that yields a zero or one probability. We only consider instances where both \storm and \mcsta can build the explicit representation of the MDP within 15 minutes.
%This was implemented by letting the tools run on default settings for all benchmarks and then making a fixed benchmark selection for the remaining experiments. 
This yields \numqvbsfull{} instances.
We obtain reference results for 344 of them from either the QVBS database or by using one of \storm's exact methods.
Reference results from different methods are always consistent.

\begin{figure}[t]
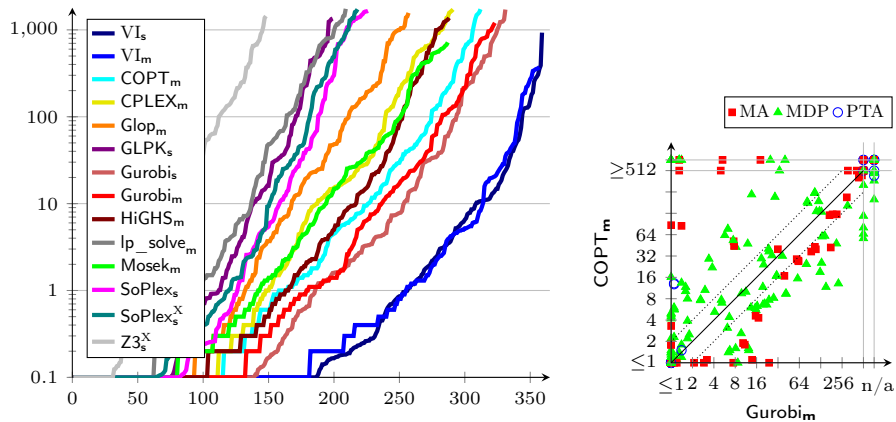

\centering
\setlength{\quantileplotwidth}{0.65\textwidth}
\setlength{\quantileplotheight}{6.5cm}
\setlength{\scatterplotsize}{0.37\textwidth}
\quantileplot{plotdata/quantile-qvbs-full.csv}
{Storm.vi/plotdarkblue, mcsta.vi/plotblue, mcsta.lp-copt-mono/plotcyan, mcsta.lp-cplex-mono/plotyellow, mcsta.lp-glop-mono/plotorange, Storm.lp-glpk-nobnds/plotpurple, Storm.lp-nobnds/plotlightred, mcsta.lp-gurobi-mono/plotred, mcsta.lp-highs-mono/plotdarkred, mcsta.lp-lpsolve-mono/plotdarkgray, mcsta.lp-mosek-mono/plotgreen, Storm.lp-soplex-nobnds/plotpink, Storm.lp-soplex-exactnobnds/plotteal, Storm.lp-z3-exactnobnds/plotlightgray}
{$\text{VI}_\tool{s}$, $\text{VI}_\tool{m}$, $\solver{COPT}_\tool{m}$, $\solver{CPLEX}_\tool{m}$, $\solver{Glop}_\tool{m}$, $\solver{GLPK}_\tool{s}$, $\solver{Gurobi}_\tool{s}$, $\solver{Gurobi}_\tool{m}$, $\solver{HiGHS}_\tool{m}$, $\solver{lp\_solve}_\tool{m}$, $\solver{Mosek}_\tool{m}$, $\solver{SoPlex}_\tool{s}$, $\solver{SoPlex}_\tool{s}^\mathrm{X}$, $\solver{Z3}_\tool{s}^\mathrm{X}$ }
{0}{365}{0.1}{1800}{north west}
\hfill
\scatterplot{plotdata/scatter-qvbs-full.csv}{mcsta.lp-gurobi-mono}{$\solver{Gurobi}_\tool{m}$}{mcsta.lp-copt-mono}{$\solver{COPT}_\tool{m}$}{true}%
\caption{Comparison of LP solver runtime on the \textit{qvbs} set}
\label{fig:LPSolvers}
\end{figure}

For LP, we have various solvers with various parameters each, cf.\ \Cref{sec:zoom-lp}. % and ways to formulate/preprocess the problem.
For conciseness, we first compare all available LP solvers on the \textit{qvbs} set.
For the best-performing solver, we then evaluate the benefit of different solver configurations.
We do the same for the choice of Markov chain solution method in PI.
We then focus on these single, reasonable, setups for LP and PI each. % compare to the other algorithms under different preprocessing steps.
%By default, we use standard floating-point implementations, and mostly separately evaluate exact ones that use arbitrary-precision rationals.

\begin{wraptable}[14]{r}{45mm}
%\begin{minipage}[b]{0.285\linewidth}
\centering
%\vspace{-2.5em}
\caption{LP summary}
\label{tab:LPSolvers}
\scriptsize
\begin{tabular}[b]{lrrr}\toprule
solver & correct & incorr. & no\,result \\\midrule
$\text{VI}_\tool{s}$ & 	359 & 	8 & 	0\\
$\text{VI}_\tool{m}$ & 	357 & 	8 & 	2\\
$\solver{COPT}_\tool{m}$ & 	312 & 	12 & 	43\\
$\solver{CPLEX}_\tool{m}$ & 	291 & 	10 & 	66\\
$\solver{Glop}_\tool{m}$ & 	257 & 	4 & 	106\\
$\solver{GLPK}_\tool{s}$ & 	199 & 	5 & 	163\\
$\solver{Gurobi}_\tool{s}$ & 	331 & 	4 & 	32\\
$\solver{Gurobi}_\tool{m}$ & 	323 & 	4 & 	40\\
$\solver{HiGHS}_\tool{m}$ & 	288 & 	10 & 	69\\
$\solver{lp\_solve}_\tool{m}$ & 	209 & 	0 & 	158\\
$\solver{Mosek}_\tool{m}$ & 	287 & 	15 & 	65\\
$\solver{SoPlex}_\tool{s}$ & 	226 & 	9 & 	132\\
$\solver{SoPlex}_\tool{s}^\mathrm{X}$ & 	218 & 	0 & 	149\\
$\solver{Z3}_\tool{s}^\mathrm{X}$  & 	148 & 	0 & 	219\\
\bottomrule
\end{tabular}
\end{wraptable}
\paragraph{LP solver comparison.}
The left-hand plot of \Cref{fig:LPSolvers} summarises the results of our comparison of the different LP solvers.
Subscript~$_\tool{s}$ and $_\tool{m}$ indicates whether solver is embedded in either \storm or \mcsta.
We apply no optimisations or reductions to the MDPs except for the precomputation of probability-0 states (and in \storm also of probability-1 states), and use the default settings for all solvers, with the trivial variable bounds $[0,1]$ and $[0, \infty)$ for probabilities and expected rewards, respectively.
%All solvers use floating-point numbers except for \solver{Z3}, which has no floating-point mode and thus computes exact (rational) results.
We include VI as baseline.
In \Cref{tab:LPSolvers}, we summarise the results.

In terms of \textbf{performance} and scalability, \solver{Gurobi} solves the highest number of benchmarks in any given time budget, closely followed by \solver{COPT}.
%(The difference between \storm and \mcsta is likely due to the differences in precomputations.)
\solver{CPLEX}, \solver{HiGHS}, and \solver{Mosek} make up a middle-class group.
%Of the floating-point-based (\emph{fp}) solvers, \solver{GLPK} and \solver{lp\_solve} are the slowest;
While the exact solver \solver{Z3} is very slow, \solver{SoPlex}'s exact mode actually competes with some fp solvers.
The quantile plots do not tell the whole story. On the right of \Cref{fig:LPSolvers}, we compare \solver{COPT} and \solver{Gurobi} which
each of them has a large number of instances on which it is (much) better.%; it is only in total that \solver{Gurobi} is somewhat faster.

In terms of \textbf{reliability} of results, the exact solvers as expected produce no incorrect results; so does the slowest fp solver, \solver{lp\_solve}.
\solver{COPT}, \solver{CPLEX}, \solver{HiGHS}, \solver{Mosek}, and fp-\solver{SoPlex} perform badly in this metric, producing more errors than VI.
Interestingly, these are mostly the faster solvers, the exception being \solver{Gurobi}.
% that is about as reliable as the much slower \solver{Glop} and \solver{GLPK}.
%The faster solvers also tend to be more quickly incorrect on our adversarial example (\cf \Cref{tab:hmmdp}).
%Possibly their default tolerances are more tuned for performance than precision.

Overall, \solver{Gurobi} achieves highest performance at decent reliability; in the remainder of this section, we thus use $\solver{Gurobi}_\tool{s}$ whenever we apply non-exact LP.

\begin{figure}[t]
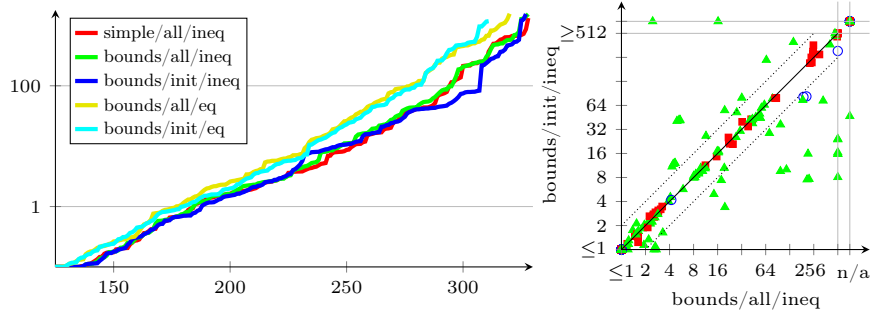

\setlength{\quantileplotwidth}{0.65\textwidth}
\setlength{\quantileplotheight}{5cm}
\setlength{\scatterplotsize}{0.4\textwidth}
\quantileplot{plotdata/quantile-qvbs-full.csv}
{Storm.lp-gurobi-4autonobnds/color1, Storm.lp-gurobi-4auto/color2, Storm.lp-gurobi-4autoinit/color3, Storm.lp-gurobi-4autoeq/color4, Storm.lp-gurobi-4autoiniteq/color5}
{simple/all/ineq, bounds/all/ineq, bounds/init/ineq, bounds/all/eq, bounds/init/eq}
{125}{330}{0.1}{1800}{north west}%
\scatterplot{plotdata/scatter-qvbs-full.csv}{Storm.lp-gurobi-4auto}{bounds/all/ineq}{Storm.lp-gurobi-4autoinit}{bounds/init/ineq}{false}
\caption{Performance impact of LP problem formulation variants (using $\solver{Gurobi}_\tool{s}$)}
\label{fig:LPTweaking}
\end{figure}

\paragraph{LP solver tweaking.}
\solver{Gurobi} can be configured to use
an ``\textit{auto}'' portfolio approach, potentially running multiple algorithms concurrently on multiple threads,
a primal or a dual simplex algorithm, or a barrier method algorithm.
We compared each option with 4 threads and found no significant performance difference.
Similarly, running the \textit{auto} method with 1, 4, and 16 threads (only here, we allocate 16  threads per experiment) also failed to bring out noticeable performance differences.
Using more threads results in a few more out-of-memory errors, though.
We thus fix \solver{Gurobi} on \textit{auto} with 4 threads.
%Quantile and scatter plots in appendix.

\Cref{fig:LPTweaking} shows the performance impact of supplying \solver{Gurobi} with more precise bounds on the variables for expected reward objectives using methods from \cite{BKLPW17,MLG05}
%DS-MPI~\cite{MLG05} for lower and the precomputations of Baier \etal\cite{BKLPW17} for upper bounds
(``bounds'' instead of ``simple''), of optimising only for initial state (``init'') instead of the sum over all states (``all''), and of using equality (``eq'') instead of less-/greater-than-or-equal (``ineq'') for unique action states.
More precise bounds yield a very small improvement at essentially no cost.
%\todo{Maxi: To me, simple looks better in Fig 4 most of the time (around 250 as well as around 300). Can we rather say that there is basically no difference? But that would change all later plots, would it?}.
Optimising for the initial state only results in a little better overall performance (in the ``pocket'' in the quantile plot around $x = 375$ that is also clearly visible in the scatter plot).
However, it also results in 2 more incorrect results in the \textit{qvbs} set. 
%We observed a similar small detrimental effect on reliability in earlier ad-hoc experiments.
Using equality for unique actions noticeably decreases performance and increases the incorrect result count by 9 instances.
For all experiments that follow, we thus use the more precise bounds, but do not enable the other two optimisations.

\begin{wrapfigure}[7]{r}{0.46\linewidth}
\vspace{-2.5em} % needed when the next paragraph is not the first on a page
\setlength{\quantileplotheight}{4cm}%
\centering%
\quantileplot{plotdata/quantile-qvbs-full.csv}
{Storm.pi/color1, Storm.pi-vi/color2, Storm.pi-ovi/color3, Storm.pi-lu/color4, Storm.pi-exactlu/color5}
{PI/gmres, PI/VI, PI/OVI, PI/LU, PI/LU$^\mathrm{X}$}
{0}{330}{0.1}{1800}{north west}
\end{wrapfigure}

\paragraph{PI methods comparison.}
The main choice in PI is which algorithm to use to solve the induced Markov chains.
On the right, we show the performance of the different algorithms available in \storm (\cf \Cref{sec:zoom-pi}).
LU$^\mathrm{X}$ yields a fully exact PI.
This interestingly performs better than the fp version, potentially because fp errors induce spurious policy changes.
The same effect likely also hinders the use of OVI, whereas VI leads to good performance.
Nevertheless, gmres is best overall, and thus our choice for all following experiments with non-exact PI.
VI and gmres yield 6 and 4 incorrect results, respectively. OVI and the exact methods are always correct on this benchmark set.

\paragraph{Best MDP algorithms for QVBS.}
We now compare all MDP model checking algorithms on the \textit{qvbs} set:
with floating-point numbers,
LP and PI configured as described above, plus unsound VI, sound OVI, and the warm-start variants of PI and LP denoted ``VI2PI'' and ``VI2LP'', respectively.
Exact results are provided by rational search (RS, essentially an exact version of VI)~\cite{DBLP:journals/fmsd/MathurBCSV20}, PI with exact LU, and LP with exact solvers (\solver{SoPlex} and \solver{Z3}).
All are implemented in \storm.

\begin{figure}[t]
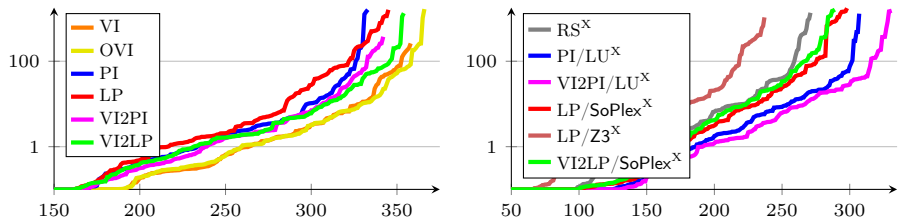

\centering
\setlength{\quantileplotwidth}{0.55\textwidth}
\setlength{\quantileplotheight}{4cm}
\quantileplot{plotdata/quantile-qvbs-full.csv}
{Storm.vi-mecq-topo/color6, Storm.ovi-topo/color4, Storm.pi-mecq-topo/color3, Storm.lp-mecq-topo-gurobi-4auto/color1, Storm.vi2pi-mecq-topo-gmres/color7, Storm.vi2lp-mecq-topo-gurobi/color2}
{VI, OVI, PI, LP, VI2PI, VI2LP}
{150}{375}{0.1}{1800}{north west}%
\quantileplot{plotdata/quantile-qvbs-full.csv}
{Storm.rs-mecq-topo-exact/color9, Storm.pi-mecq-topo-exactlu/color3, Storm.vi2pi-mecq-topo-exactlu/color7, Storm.lp-mecq-topo-soplex-exact/color1, Storm.lp-mecq-topo-z3-exact/color16, Storm.vi2lp-mecq-topo-soplex-exact/color2}
{RS$^\mathrm{X}$, PI/LU$^\mathrm{X}$, VI2PI/LU$^\mathrm{X}$, LP/$\solver{SoPlex}^\mathrm{X}$, LP/$\solver{Z3}^\mathrm{X}$, VI2LP/$\solver{SoPlex}^\mathrm{X}$}
{50}{335}{0.1}{1800}{north west}
\caption{Comparison of MDP model checking algorithms on the \textit{qvbs} set}
\label{fig:MethodsQVBS}
\end{figure}

In a first experiment, we evaluated the impact of using the topological approach and of collapsing MECs (\cf \Cref{sec:prelim-opts}).
The results, for which we omit plots, are that the topological approach noticeably improves performance and scalability for \emph{all} algorithms, and we therefore always use it from now on.
Collapsing MECs is necessary to guarantee termination of OVI, while for the other algorithms it is a potential optimisation; however we found it to overall have a minimal positive performance impact only.
Since it is required by OVI and does not reduce performance, we also always use it from now on.

\begin{figure}[t]
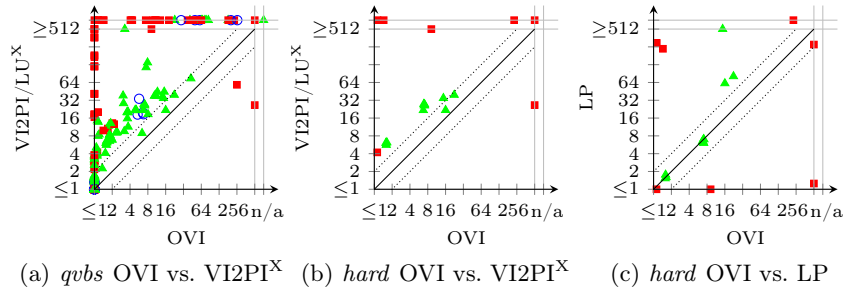

	\setlength{\scatterplotsize}{0.33\textwidth}
    \centering
    \subfigure[\emph{qvbs} OVI vs.\ VI2PI$^\mathrm{X}$]{
    \scatterplot{plotdata/scatter-qvbs-full.csv}{Storm.ovi-topo}{OVI}{Storm.vi2pi-mecq-topo-exactlu}{VI2PI/LU$^\mathrm{X}$}{false} \label{fig:qvbsovipi}}
	\hspace{-2em}
	\subfigure[\emph{hard} OVI vs.\ VI2PI$^\mathrm{X}$]{
		\scatterplot{plotdata/scatter-qvbs-hard.csv}{Storm.ovi-topo}{OVI}{Storm.vi2pi-mecq-topo-exactlu}{VI2PI/LU$^\mathrm{X}$}{false} \label{fig:hardbot}
	}
	\hspace{-2em}
    \subfigure[\emph{hard} OVI vs.\ LP]{
    \scatterplot{plotdata/scatter-qvbs-hard.csv}{Storm.ovi-topo}{OVI}{Storm.lp-mecq-topo-gurobi-4auto}{LP}{false} 
    \label{fig:hardtop}
    }
    
     \vspace{-1em}
    \caption{Additional direct performance comparisons}
    \label{fig:my_label}
   
\end{figure}

\Cref{fig:MethodsQVBS} shows the complete comparison of all the methods on the \textit{qvbs} set, for fp algorithms on the left and exact solutions on the right.
Among the fp algorithms, OVI is clearly the fastest and most scalable.
VI is somewhat faster but incurs several incorrect results that diminish its appearance in the
%\begin{wrapfigure}[9]{r}{0.325\linewidth}
%\vspace{-2.25em} % needed when the next paragraph is not the first on a page
%\centering
%\scatterplot{plotdata/scatter-qvbs-full.csv}{Storm.ovi-topo}{OVI}{Storm.vi2pi-mecq-topo-exactlu}{VI2PI/LU$^\mathrm{X}$}{false}
%\end{wrapfigure}
quantile plot.
OVI is additionally special among these algorithms in that it is sound, \ie provides guaranteed $\epsilon$-correct results---though up to fp rounding errors, which can be eliminated following the approach of~\cite{H22}.
On the exact side, PI with an inexact-VI warm start works best.
The scatter plot in Fig.~\ref{fig:qvbsovipi} shows the performance impact of computing an exact instead of an approximate solution.

\begin{figure}[t]
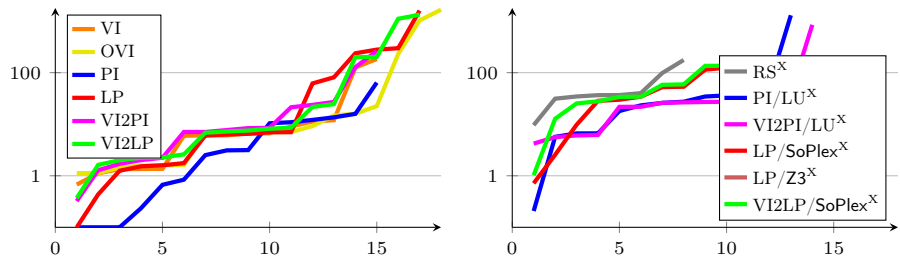

\centering
\setlength{\quantileplotwidth}{0.55\textwidth}
\setlength{\quantileplotheight}{4.5cm}
\quantileplot{plotdata/quantile-qvbs-hard.csv}
{Storm.vi-mecq-topo/color6, Storm.ovi-topo/color4, Storm.pi-mecq-topo/color3, Storm.lp-mecq-topo-gurobi-4auto/color1, Storm.vi2pi-mecq-topo-gmres/color7, Storm.vi2lp-mecq-topo-gurobi/color2}
{VI, OVI, PI, LP, VI2PI, VI2LP}
{0}{\numqvbshard}{0.1}{1800}{north west}%
\quantileplot{plotdata/quantile-qvbs-hard.csv}
{Storm.rs-mecq-topo-exact/color9, Storm.pi-mecq-topo-exactlu/color3, Storm.vi2pi-mecq-topo-exactlu/color7, Storm.lp-mecq-topo-soplex-exact/color1, Storm.lp-mecq-topo-z3-exact/color16, Storm.vi2lp-mecq-topo-soplex-exact/color2}
{RS$^\mathrm{X}$, PI/LU$^\mathrm{X}$, VI2PI/LU$^\mathrm{X}$, LP/$\solver{SoPlex}^\mathrm{X}$, LP/$\solver{Z3}^\mathrm{X}$, VI2LP/$\solver{SoPlex}^\mathrm{X}$}
{0}{\numqvbshard}{0.1}{1800}{south east}
\caption{Comparison of MDP model checking algorithms on the \textit{hard} subset}
\label{fig:MethodsHard}
\end{figure}

\subsection{The Hard QVBS Benchmarks}\label{sec:exp-hard}

The QVBS contains many models built for tools that use VI as default algorithm.
The other algorithms may actually be important to solve key challenging instances where VI/OVI perform badly.
This contribution could be hidden in the sea of instances trivial for VI.
We thus zoom in on a selection of QVBS instances that appear ``hard'' for VI:
those where VI takes longer than the prior MDP state space construction phase in both \storm and \mcsta, and additionally both phases together take at least 1\,s.
These are \numqvbshard{} of the previously considered \numqvbsfull{} instances.

%\begin{wrapfigure}[18]{r}{0.325\linewidth}
%\vspace{-2.25em} % needed when the next paragraph is not the first on a page
%\centering
%\scatterplot{plotdata/scatter-qvbs-hard.csv}{Storm.ovi-topo}{OVI}{Storm.lp-mecq-topo-gurobi-4auto}{LP}{false}\\
%\scatterplot{plotdata/scatter-qvbs-hard.csv}{Storm.ovi-topo}{OVI}{Storm.vi2pi-mecq-topo-exactlu}{VI2PI/LU$^\mathrm{X}$}{false}
%\end{wrapfigure}

In \Cref{fig:MethodsHard}, we show the behaviour of all the algorithms on this \textit{hard} subset.
OVI again works better than VI due to the incorrect results that VI returns.
We see that the performance and scalability gap between the algorithms has narrowed; although OVI still ``wins'', LP in particular is much closer than on the full \textit{qvbs} set.
We also investigated the LP outcomes with solvers other than \solver{Gurobi}:
even on this set, \solver{Gurobi} and \solver{COPT} remain the fastest and most scalable solvers.
With \mcsta, in the basic configuration, they solve 16 and 17 instances, the slowest taking 835\,s and 1334\,s, respectively;
with topo, the numbers become 17 and 15 instances with the slowest at 1373\,s and 1590\,s seconds.
%\todo{I think we don't have space for these details?}
We show the detailed comparison of OVI and LP, noting that there are a few instances where LP is much faster (Fig.~\ref{fig:hardtop}), and
%Here the scatterplot for OVI vs.\ LP: a few instances where LP is actually much faster (we note that these are most of the ones where it is faster on full, too, \ie full doesn't have many more where LP is better).
repeat the comparison between the best fp and exact algorithms (Fig.~\ref{fig:hardbot}).

\begin{figure}[t]
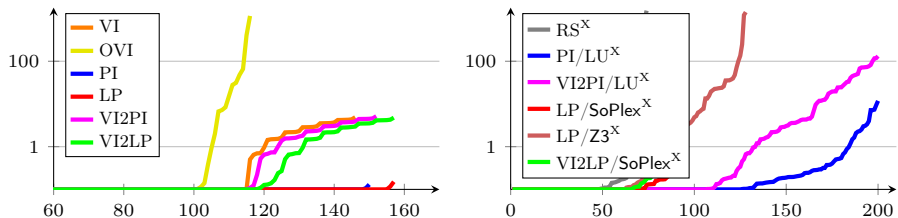

\centering
\setlength{\quantileplotwidth}{0.55\textwidth}
\setlength{\quantileplotheight}{4cm}
\quantileplot{plotdata/quantile-premise.csv}
{Storm.vi-mecq-topo/color6, Storm.ovi-topo/color4, Storm.pi-mecq-topo/color3, Storm.lp-mecq-topo-gurobi-4auto/color1, Storm.vi2pi-mecq-topo-gmres/color7, Storm.vi2lp-mecq-topo-gurobi/color2}
{VI, OVI, PI, LP, VI2PI, VI2LP}
{60}{170}{0.1}{1800}{north west}%
\quantileplot{plotdata/quantile-premise.csv}
{Storm.rs-mecq-topo-exact/color9, Storm.pi-mecq-topo-exactlu/color3, Storm.vi2pi-mecq-topo-exactlu/color7, Storm.lp-mecq-topo-soplex-exact/color1, Storm.lp-mecq-topo-z3-exact/color16, Storm.vi2lp-mecq-topo-soplex-exact/color2}
{RS$^\mathrm{X}$, PI/LU$^\mathrm{X}$, VI2PI/LU$^\mathrm{X}$, LP/$\solver{SoPlex}^\mathrm{X}$, LP/$\solver{Z3}^\mathrm{X}$, VI2LP/$\solver{SoPlex}^\mathrm{X}$}
{0}{210}{0.1}{1800}{north west}
\caption{Comparison of MDP model checking algorithms on the \textit{premise} set}
\label{fig:MethodsPremise}
\end{figure}

\subsection{The Runtime Monitoring Benchmarks}
\label{sec:runtimemonitors} 

% Genau, das sind alles runtime monitoring probleme wo wir A) stochastic noise haben, B) unobservable and uncontrollable nondeterminism und C) partial state observations. Der Trace da oben ist dann die observierte history

% Ok, erst informell: Das ist alles conditional probabilities mit einen Baier-et-al unrolling
% dh wir haben einen MDP und einen trace der laenge N, N=50 bis 1000. Jetzt falten wir den MDP also N mal aus. Wenn das label an der stelle uebereinstimmt gehen wir von da in episode N+1, ansonsten resetten wir zum starting state

% und die Wkeit die wir berechnen ist eine normalisiertes Risiko um das System weiter laufen zu lassen
% und das Problem wird dann intern in ein model checking problem uebersetzt

While the QVBS is intentionally diverse, our third set of benchmarks is intentionally focused:
We study \numpremise{} MDPs from a runtime monitoring study~\cite{DBLP:conf/cav/JungesTS20}.
The original problem is to compute the normalised risk of continuing to operate the system being monitored subject to stochastic noise, unobservable and uncontrollable nondeterminism, and partial state observations.
This is a query for a conditional probability.
It is answered via probabilistic model checking by unrolling an MDP model model along an observed history trace of length $n \in \set{50, \ldots, 1000}$ following the approach of Baier \etal\cite{BKKM14}.
The MDPs contain many transitions back to the initial state, ultimately resulting in numerically challenging instances (compare the structure of $M_n$ in \Cref{sec:prelim-guarantees}).
We were able to compute a reference result for all instances.

\Cref{fig:MethodsPremise} compares the different MDP model checking algorithms on this set.
In line with the observations in~\cite{DBLP:conf/cav/JungesTS20}, we see very different behaviour compared to the QVBS.
Among the fp solutions on the left, LP with \solver{Gurobi} terminates very quickly (under $1$\,s), and either produces a correct (155 instances) or a completely incorrect result (mostly $0$, on 45 instances).
VI behaves similarly, but is slower.
OVI, in contrast, delivers no incorrect result, but instead fails to terminate on all but 116 instances.
In the exact setting, warm starts using VI inherit its relative slowness and consequently do not pay off.
Exact PI outperforms both exact LP solvers.
In the case of exact \solver{SoPlex}, out of the 112 instances it does not manage to solve, 98 are errors likely related to a confirmed bug in the current version.
%\footnote{See
% the github issue at
%\href{https://github.com/scipopt/soplex/issues/7}{github.com/scipopt/soplex/issues/7}.
%\textit{(link removed for double-blind reviewing)}.
%}.

The \textit{premise} set highlights that the best MDP model checking algorithm depends on the application. % application determines which the best MDP model checking algorithm is.
Here, in the fp case, LP appears best but produces unreliable (incorrect) results; the seemingly much worse OVI at least does not do so.
Given the numeric challenge, an exact method should be chosen, and we show that these actually perform well here.

\iffalse
% Integrated in conclusion
\subsection{Beyond Our Benchmark Sets}

We were unable to find further sets of MDPs that are hard for VI. 
Several stochastic games (SGs) difficult for VI were found in~\cite{KRSW22}; the authors noted that using PI for the SGs %and then LP for the induced MDPs - in fact, there we still used PRISM solution on the Markov chain, which was only VI. Only the Maxi-ATVA paper considers LP.
was better than applying VI to the SGs.
However, when we extracted the induced MDPs, we found them all easy for VI.
Similarly, \cite{Maxi-ATVA-arxiv} used a random generation of SGs of at most 10,000 states, many of which were challenging for the SG algorithms.
Yet the same random generation modified to produce MDPs delivered only MDPs easily solved in seconds, even with drastically increased numbers of states.
%Even scaling up the models to the point where building them took 20 minutes and reducing the occurring transition probabilities did not yield a challenging model.
This is in contrast to what Alag{\"{o}}z \etal\cite{AAL15} found:
their random generation returned models where LP beat PI. 
However, their setting is discounted, and their description of the random generation was too superficial for us to be able to replicate it. % reward and they don't describe their random generation precisely enough such that one could expect to reproduce their results. 
\fi

%% file: 9_conclusion.tex
\section{Conclusion}

We thoroughly investigated the state of the art in MDP model checking, showing that there is no single best algorithm for this task. 
For benchmarks which are not numerically challenging, OVI is a sensible default, closely followed by PI and LP with a warm start---although using the latter two means losing soundness as confirmed by a number of incorrect results in our experiments.
For numerically hard benchmarks, PI and LP as well as computing exact solutions are more attractive, and clearly preferable in combination.
Overall, although LP has the superior (polynomial) theoretical complexity, it in our practical evaluation almost always performs worse than the other (exponential) approaches.
This is even though we use modern commercial solvers and tune both the LP encoding of the problem as well as the solvers' parameters.
%
%Ample future work remains:
While we \emph{observed} the behaviour of the different algorithms and have some intuition into what makes the \textit{premise} set hard, an entire research questions of its own is to identify and quantify the structural properties that make a model hard.

Our evaluation also raises the question of how prevalent MDPs that challenge VI are in practice.
Aside from the \textit{premise} benchmarks, we were unable to find further sets of MDPs that are hard for VI. 
Notably, several stochastic games (SGs) difficult for VI were found in~\cite{KRSW22}; the authors noted that using PI for the SGs %and then LP for the induced MDPs - in fact, there we still used PRISM solution on the Markov chain, which was only VI. Only the Maxi-ATVA paper considers LP.
was better than applying VI to the SGs.
However, when we extracted the induced MDPs, we found them all easy for VI.
Similarly, \cite{Maxi-ATVA-arxiv} used a random generation of SGs of at most 10,000 states, many of which were challenging for the SG algorithms.
Yet the same random generation modified to produce MDPs delivered only MDPs easily solved in seconds, even with drastically increased numbers of states.
%Even scaling up the models to the point where building them took 20 minutes and reducing the occurring transition probabilities did not yield a challenging model.
In contrast, Alag{\"{o}}z \etal\cite{AAL15} report that their random generation returned models where LP beat PI.
However, their setting is discounted, and their description of the random generation was too superficial for us to be able to replicate it. % reward and they don't describe their random generation precisely enough such that one could expect to reproduce their results.
We note that, in several of our scatter plots, the MA instances from the QVBS (where we check the embedded MDP) appeared more challenging overall than the MDPs.
We thus conclude this paper with a call for challenging MDP benchmarks---as separate benchmark sets of unique characteristics like \textit{premise}, or for inclusion in the QVBS.

%More properties; 
%An experimental comparison for long-run averages~\cite{ACD+17} showed that LP indeed did not scale well to models with many maximal ECs.
%
%In the loop solving
%
%An obvious questions is what causes models to be hard and analyze structure. This is a hard problem, see e.g. ATVA-Maxi, which didn't manage to do that.
%
%TODO-list for later:
%\begin{itemize}
%    \item mention thresholds, maybe on a few examples in detail.
%“Opens up new question like …”
%[@Journal/separate paper: explain how thresholds affect (unsat -> only upper VI), but needs to compare to Moritz’ tool, partial exploration…]
%    \item There is more stuff to try for LP, but it does not look promising, since we tried a lot and still didn't become competitive. 
%    VI2LP, giro method (setting base) vs. just setting the bounds
%    \item Trying more for PI seems reasonable. Practically, maybe there are cool things like SVI and OVI. Also: Implementation of modified PI.
%    \item LP investigate variability with random seed (MO of mcsta and Storm are different. Give explicit model with fixed state ordering and see whether it makes a difference). See Signal 13.10. ~14:00
%    \item Set tolerances explicitly in the different solvers and check their influence.
%\end{itemize}

%% file: appendix.tex
\appendix
\newpage
%\section{Additional plots}

\section{The QVBS Benchmarks: LP solver tweaking}\label{app:QVBS-LP-tweaking}

\begin{figure}[h]
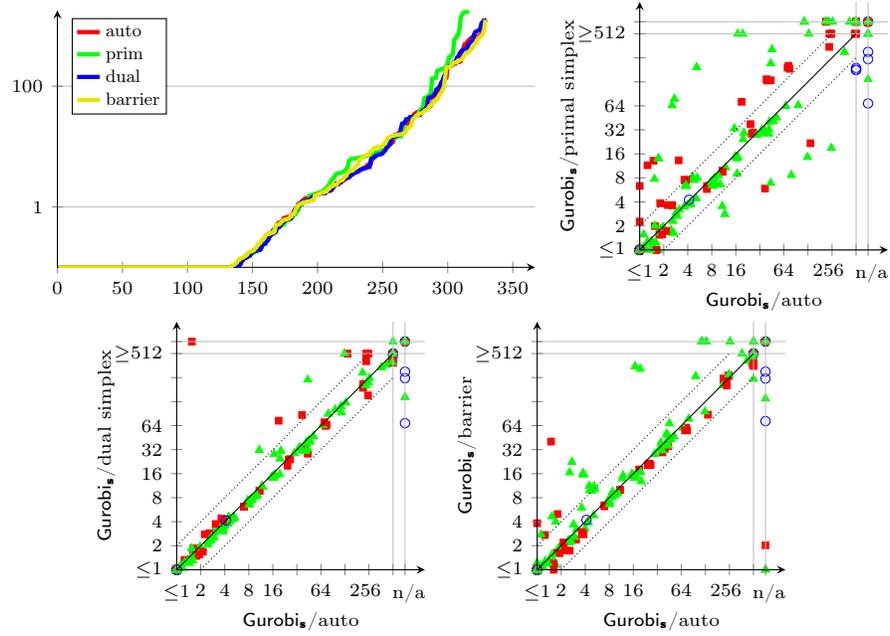

	\centering
	\setlength{\quantileplotwidth}{0.65\textwidth}
	\setlength{\quantileplotheight}{5cm}
	\setlength{\scatterplotsize}{0.4\textwidth}
	\quantileplot{plotdata/quantile-qvbs-full.csv}
	{Storm.lp-gurobi-4autonobnds/color1, Storm.lp-gurobi-4primalsimplnobnds/color2, Storm.lp-gurobi-4dualsimplnobnds/color3, Storm.lp-gurobi-4barriernobnds/color4}
	{auto, prim, dual, barrier}
	{0}{\numqvbsfull}{0.1}{1800}{north west}
	\hfill
	\scatterplot{plotdata/scatter-qvbs-full.csv}{Storm.lp-gurobi-4autonobnds}{$\solver{Gurobi}_\tool{s}$/auto}{Storm.lp-gurobi-4primalsimplnobnds}{$\solver{Gurobi}_\tool{s}$/primal simplex}{false}
	
	\scatterplot{plotdata/scatter-qvbs-full.csv}{Storm.lp-gurobi-4autonobnds}{$\solver{Gurobi}_\tool{s}$/auto}{Storm.lp-gurobi-4dualsimplnobnds}{$\solver{Gurobi}_\tool{s}$/dual simplex}{false}
	\scatterplot{plotdata/scatter-qvbs-full.csv}{Storm.lp-gurobi-4autonobnds}{$\solver{Gurobi}_\tool{s}$/auto}{Storm.lp-gurobi-4barriernobnds}{$\solver{Gurobi}_\tool{s}$/barrier}{false}
	\caption{Comparison of \solver{Gurobi}'s configurations. See the paragraph \emph{LP solver tweaking} in \cref{sec:exp-qvbs} for a discussion of the results.}
\end{figure}

\begin{figure}[h]
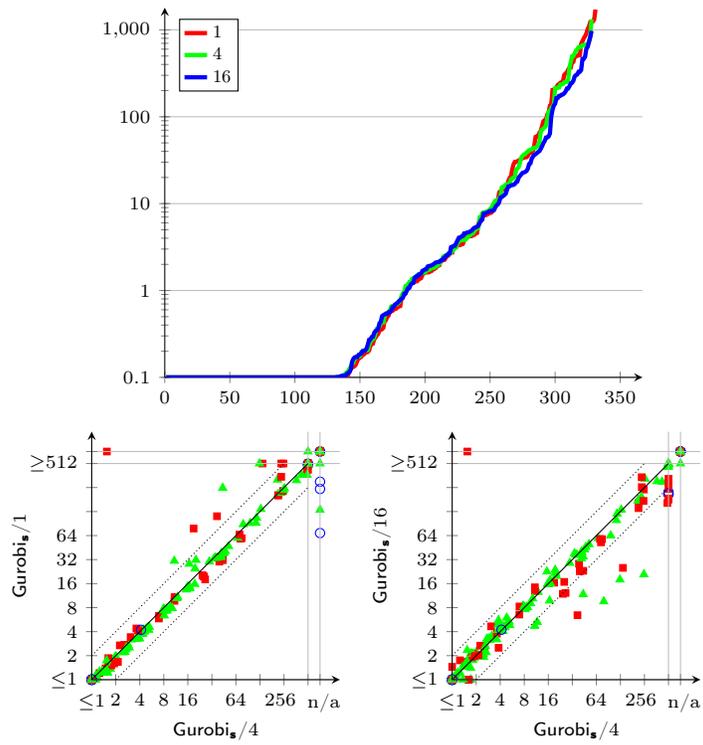

	\centering
	\setlength{\quantileplotwidth}{0.65\textwidth}
	\setlength{\quantileplotheight}{6.5cm}
	\setlength{\scatterplotsize}{0.4\textwidth}
	\quantileplot{plotdata/quantile-qvbs-full.csv}
	{Storm.lp-nobnds/color1, Storm.lp-gurobi-4autonobnds/color2, Storm.lp-gurobi-16autonobnds/color3}
	{1, 4, 16}
	{0}{\numqvbsfull}{0.1}{1800}{north west}

\scatterplot{plotdata/scatter-qvbs-full.csv}{Storm.lp-gurobi-4autonobnds}{$\solver{Gurobi}_\tool{s}$/4}{Storm.lp-nobnds}{$\solver{Gurobi}_\tool{s}$/1}{false}
\scatterplot{plotdata/scatter-qvbs-full.csv}{Storm.lp-gurobi-4autonobnds}{$\solver{Gurobi}_\tool{s}$/4}{Storm.lp-gurobi-16autonobnds}{$\solver{Gurobi}_\tool{s}$/16}{false}
	\caption{Comparison of how the number of threads affect the performance of \solver{Gurobi}'s auto method. See the paragraph \emph{LP solver tweaking} in \cref{sec:exp-qvbs} for a discussion of the results.}
\end{figure}

\begin{figure}[h]
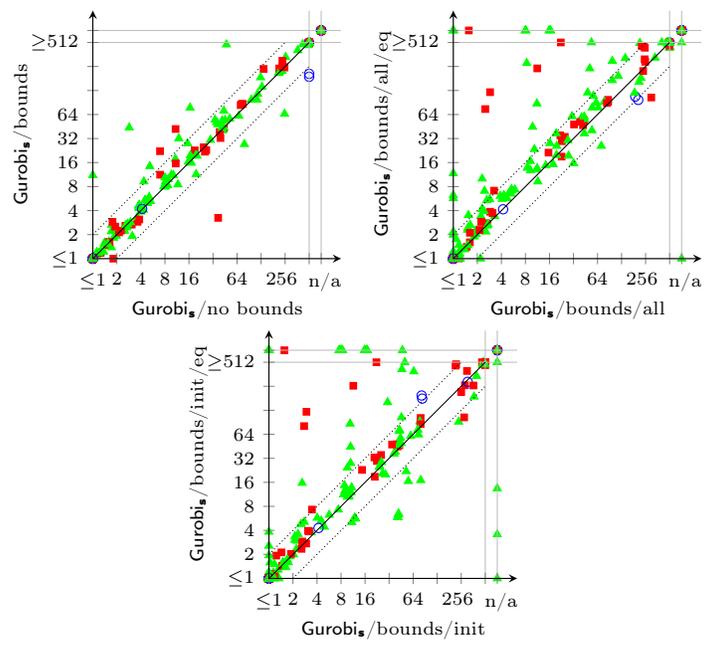

	\centering
	\setlength{\scatterplotsize}{0.4\textwidth}
	\scatterplot{plotdata/scatter-qvbs-full.csv}{Storm.lp-gurobi-4autonobnds}{$\solver{Gurobi}_\tool{s}$/no bounds}{Storm.lp-gurobi-4auto}{$\solver{Gurobi}_\tool{s}$/bounds}{false}
	\scatterplot{plotdata/scatter-qvbs-full.csv}{Storm.lp-gurobi-4auto}{$\solver{Gurobi}_\tool{s}$/bounds/all}{Storm.lp-gurobi-4autoeq}{$\solver{Gurobi}_\tool{s}$/bounds/all/eq}{false}
	\scatterplot{plotdata/scatter-qvbs-full.csv}{Storm.lp-gurobi-4autoinit}{$\solver{Gurobi}_\tool{s}$/bounds/init}{Storm.lp-gurobi-4autoiniteq}{$\solver{Gurobi}_\tool{s}$/bounds/init/eq}{false}
	
	\caption{Comparison of further LP problem formulation variants. See the paragraph describing \cref{fig:LPTweaking} in \cref{sec:exp-qvbs} for a description.}
\end{figure}

\clearpage

%Variable bounds

%\scatterplot{plotdata/scatter-qvbs-full.csv}{Storm.lp-gurobi-4autonobnds}{$\solver{Gurobi}_\tool{s}$/no bounds}{Storm.lp-gurobi-4auto}{$\solver{Gurobi}_\tool{s}$/bounds}{false}

%Other tweaking

%\scatterplot{plotdata/scatter-qvbs-full.csv}{Storm.lp-gurobi-4auto}{$\solver{Gurobi}_\tool{s}$/bounds/all}{Storm.lp-gurobi-4autoeq}{$\solver{Gurobi}_\tool{s}$/bounds/all/eq}{false}
%\scatterplot{plotdata/scatter-qvbs-full.csv}{Storm.lp-gurobi-4autoinit}{$\solver{Gurobi}_\tool{s}$/bounds/init}{Storm.lp-gurobi-4autoiniteq}{$\solver{Gurobi}_\tool{s}$/bounds/init/eq}{false}

\section{QVBS Benchmarks: MEC collapsing and topological decomposition}\label{app:QVBS-topo-mecs}

\begin{figure}[h]
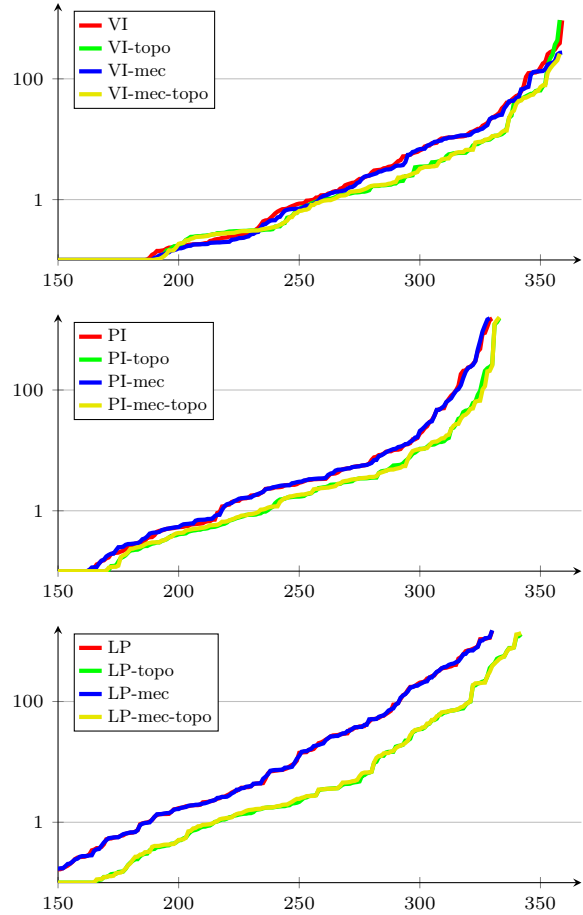

	\centering
	{\setlength{\quantileplotwidth}{0.7\textwidth}
		\setlength{\quantileplotheight}{5cm}
		\quantileplot{plotdata/quantile-qvbs-full.csv}
		{Storm.vi/color1, Storm.vi-topo/color2, Storm.vi-mecq/color3, Storm.vi-mecq-topo/color4}
		{VI, VI-topo, VI-mec, VI-mec-topo}
		{150}{\numqvbsfull}{0.1}{1800}{north west}%
	}
	
	{\setlength{\quantileplotwidth}{0.7\textwidth}
		\setlength{\quantileplotheight}{5cm}
		\quantileplot{plotdata/quantile-qvbs-full.csv}
		{Storm.pi/color1, Storm.pi-topo/color2, Storm.pi-mecq/color3, Storm.pi-mecq-topo/color4}
		{PI, PI-topo, PI-mec, PI-mec-topo}
		{150}{\numqvbsfull}{0.1}{1800}{north west}%
	}
	
	{\setlength{\quantileplotwidth}{0.7\textwidth}
		\setlength{\quantileplotheight}{5cm}
		\quantileplot{plotdata/quantile-qvbs-full.csv}
		{Storm.lp/color1, Storm.lp-topo/color2, Storm.lp-mecq/color3, Storm.lp-mecq-topo/color4}
		{LP, LP-topo, LP-mec, LP-mec-topo}
		{150}{\numqvbsfull}{0.1}{1800}{north west}%
	}
	\caption{Comparison of the vanilla algorithms VI, PI and LP and their variants using MEC collapsing and topological decomposition. See the discussion below the paragraph \emph{Best MDP algorithms for QVBS} in \cref{sec:exp-qvbs} for a discussion of the results.}
\end{figure}

%\scatterplot{plotdata/scatter-qvbs-full.csv}{Storm.vi}{VI}{Storm.vi-topo}{VI-topo}{false}
%\scatterplot{plotdata/scatter-qvbs-full.csv}{Storm.vi}{VI}{Storm.vi-mecq}{VI-mecq}{false}
%\scatterplot{plotdata/scatter-qvbs-full.csv}{Storm.vi}{VI}{Storm.vi-mecq-topo}{VI-mecq-topo}{false}
%\scatterplot{plotdata/scatter-qvbs-full.csv}{Storm.vi-topo}{VI-topo}{Storm.vi-mecq-topo}{VI-mecq-topo}{false}

\clearpage

\section{Hard benchmarks: LP solver runtime}\label{app:hard-all}

\begin{figure}[h]
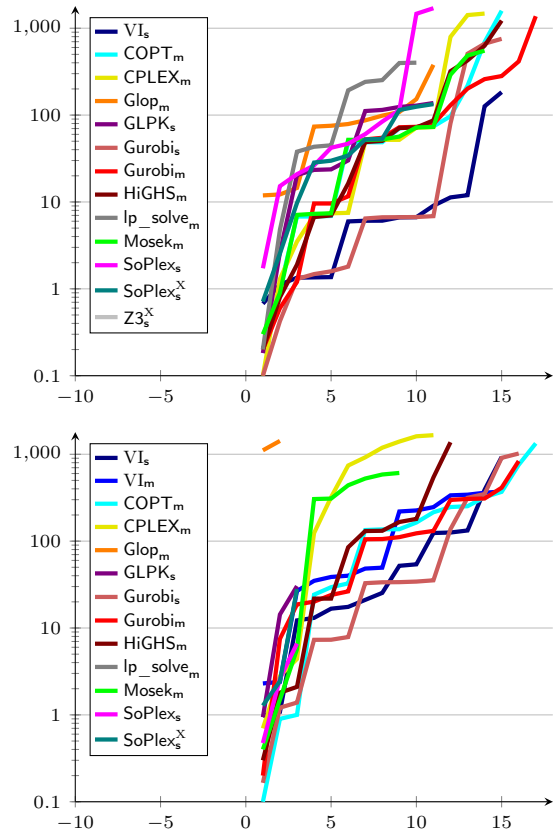

	\centering

	{
		\setlength{\quantileplotwidth}{0.65\textwidth}
		\setlength{\quantileplotheight}{6.5cm}
		\setlength{\scatterplotsize}{0.4\textwidth}
		\quantileplot{plotdata/quantile-qvbs-hard.csv}
		{Storm.vi-mecq-topo/plotdarkblue, mcsta.lp-copt-topo/plotcyan, mcsta.lp-cplex-topo/plotyellow, mcsta.lp-glop-topo/plotorange, Storm.lp-mecq-topo-glpk/plotpurple, Storm.lp-mecq-topo/plotlightred, mcsta.lp-gurobi-topo/plotred, mcsta.lp-highs-topo/plotdarkred, mcsta.lp-lpsolve-topo/plotdarkgray, mcsta.lp-mosek-topo/plotgreen, Storm.lp-mecq-topo-soplex/plotpink, Storm.lp-mecq-topo-soplex-exact/plotteal, Storm.lp-mecq-topo-z3-exact/plotlightgray}
		{$\text{VI}_\tool{s}$, $\solver{COPT}_\tool{m}$, $\solver{CPLEX}_\tool{m}$, $\solver{Glop}_\tool{m}$, $\solver{GLPK}_\tool{s}$, $\solver{Gurobi}_\tool{s}$, $\solver{Gurobi}_\tool{m}$, $\solver{HiGHS}_\tool{m}$, $\solver{lp\_solve}_\tool{m}$, $\solver{Mosek}_\tool{m}$, $\solver{SoPlex}_\tool{s}$, $\solver{SoPlex}_\tool{s}^\mathrm{X}$, $\solver{Z3}_\tool{s}^\mathrm{X}$}
		{-10}{\numqvbshard}{0.1}{1800}{north west}
	}
	
	{
		\setlength{\quantileplotwidth}{0.65\textwidth}
		\setlength{\quantileplotheight}{6.5cm}
		\setlength{\scatterplotsize}{0.4\textwidth}
		\quantileplot{plotdata/quantile-qvbs-hard.csv}
		{Storm.vi/plotdarkblue, mcsta.vi/plotblue, mcsta.lp-copt-mono/plotcyan, mcsta.lp-cplex-mono/plotyellow, mcsta.lp-glop-mono/plotorange, Storm.lp-glpk-nobnds/plotpurple, Storm.lp-nobnds/plotlightred, mcsta.lp-gurobi-mono/plotred, mcsta.lp-highs-mono/plotdarkred, mcsta.lp-lpsolve-mono/plotdarkgray, mcsta.lp-mosek-mono/plotgreen, Storm.lp-soplex-nobnds/plotpink, Storm.lp-soplex-exactnobnds/plotteal, Storm.lp-z3-exactnobnds/plotlightgray}
		{$\text{VI}_\tool{s}$, $\text{VI}_\tool{m}$, $\solver{COPT}_\tool{m}$, $\solver{CPLEX}_\tool{m}$, $\solver{Glop}_\tool{m}$, $\solver{GLPK}_\tool{s}$, $\solver{Gurobi}_\tool{s}$, $\solver{Gurobi}_\tool{m}$, $\solver{HiGHS}_\tool{m}$, $\solver{lp\_solve}_\tool{m}$, $\solver{Mosek}_\tool{m}$, $\solver{SoPlex}_\tool{s}$, $\solver{SoPlex}_\tool{s}^\mathrm{X}$, $\solver{Z3}_\tool{s}^\mathrm{X}$}
		{-10}{\numqvbshard}{0.1}{1800}{north west}
	}
	\caption{Comparison of LP solver runtime on the \emph{hard} subset. The upper plot uses the topological optimization, the lower does not. See \cref{sec:exp-hard} for a discussion of the results.}
\end{figure}

%% file: 0_main.bbl
\begin{thebibliography}{10}
\providecommand{\url}[1]{\texttt{#1}}
\providecommand{\urlprefix}{URL }
\providecommand{\doi}[1]{https://doi.org/#1}

\bibitem{AAL15}
Alag{\"{o}}z, O., Ayvaci, M.U.S., Linderoth, J.T.: Optimally solving {M}arkov
  decision processes with total expected discounted reward function: Linear
  programming revisited. Comput. Ind. Eng.  \textbf{87},  311--316 (2015).
  \doi{10.1016/j.cie.2015.05.031}

\bibitem{anand2017comparative}
Anand, R., Aggarwal, D., Kumar, V.: A comparative analysis of optimization
  solvers. Journal of Statistics and Management Systems  \textbf{20}(4),
  623--635 (2017). \doi{10.1080/09720510.2017.1395182}

\bibitem{Maxi-ATVA-arxiv}
Azeem, M., Evangelidis, A., Kret{\'{\i}}nsk{\'{y}}, J., Slivinskiy, A.,
  Weininger, M.: Optimistic and topological value iteration for simple
  stochastic games. CoRR  \textbf{abs/2207.14417} (2022).
  \doi{10.48550/arXiv.2207.14417},
  \url{https://doi.org/10.48550/arXiv.2207.14417}

\bibitem{DBLP:reference/mc/BaierAFK18}
Baier, C., de~Alfaro, L., Forejt, V., Kwiatkowska, M.: Model checking
  probabilistic systems. In: Handbook of Model Checking, pp. 963--999. Springer
  (2018)

\bibitem{DBLP:series/lncs/BaierHK19}
Baier, C., Hermanns, H., Katoen, J.: The 10, 000 facets of {MDP} model
  checking. In: Computing and Software Science, {LNCS}, vol. 10000, pp.
  420--451. Springer (2019)

\bibitem{BK08}
Baier, C., Katoen, J.: Principles of model checking. {MIT} Press (2008),
  \url{https://mitpress.mit.edu/books/principles-model-checking}

\bibitem{BKKM14}
Baier, C., Klein, J., Kl{\"{u}}ppelholz, S., M{\"{a}}rcker, S.: Computing
  conditional probabilities in {M}arkovian models efficiently. In: {TACAS}.
  {LNCS}, vol.~8413, pp. 515--530. Springer (2014).
  \doi{10.1007/978-3-642-54862-8\_43}

\bibitem{BKLPW17}
Baier, C., Klein, J., Leuschner, L., Parker, D., Wunderlich, S.: Ensuring the
  reliability of your model checker: Interval iteration for {M}arkov decision
  processes. In: {CAV} {(1)}. {LNCS}, vol. 10426, pp. 160--180. Springer
  (2017). \doi{10.1007/978-3-319-63387-9\_8},
  \url{https://doi.org/10.1007/978-3-319-63387-9\_8}

\bibitem{BK0PS19-complexityVI}
Balaji, N., Kiefer, S., Novotn{\'{y}}, P., P{\'{e}}rez, G.A., Shirmohammadi,
  M.: On the complexity of value iteration. In: {ICALP}. LIPIcs, vol.~132, pp.
  102:1--102:15. Schloss Dagstuhl - Leibniz-Zentrum f{\"{u}}r Informatik
  (2019). \doi{10.4230/LIPIcs.ICALP.2019.102},
  \url{https://doi.org/10.4230/LIPIcs.ICALP.2019.102}

\bibitem{DBLP:journals/mor/BertsekasT91}
Bertsekas, D.P., Tsitsiklis, J.N.: An analysis of stochastic shortest path
  problems. Math. Oper. Res.  \textbf{16}(3),  580--595 (1991).
  \doi{10.1287/moor.16.3.580}, \url{https://doi.org/10.1287/moor.16.3.580}

\bibitem{DBLP:books/cu/BV2014}
Boyd, S.P., Vandenberghe, L.: Convex Optimization. Cambridge University Press
  (2014)

\bibitem{BCC+14}
Br{\'{a}}zdil, T., Chatterjee, K., Chmelik, M., Forejt, V.,
  Kret{\'{\i}}nsk{\'{y}}, J., Kwiatkowska, M.Z., Parker, D., Ujma, M.:
  Verification of {M}arkov decision processes using learning algorithms. In:
  {ATVA}. {LNCS}, vol.~8837, pp. 98--114. Springer (2014).
  \doi{10.1007/978-3-319-11936-6\_8},
  \url{https://doi.org/10.1007/978-3-319-11936-6\_8}

\bibitem{DBLP:conf/isola/BuddeHKKPQTZ20}
Budde, C.E., Hartmanns, A., Klauck, M., Kret{\'{\i}}nsk{\'{y}}, J., Parker, D.,
  Quatmann, T., Turrini, A., Zhang, Z.: On correctness, precision, and
  performance in quantitative verification -- {QC}omp 2020 competition report.
  In: ISoLA {(4)}. {LNCS}, vol. 12479, pp. 216--241. Springer (2020)

\bibitem{CH08}
Chatterjee, K., Henzinger, T.A.: Value iteration. In: 25 Years of Model
  Checking. {LNCS}, vol.~5000, pp. 107--138. Springer (2008).
  \doi{10.1007/978-3-540-69850-0\_7},
  \url{https://doi.org/10.1007/978-3-540-69850-0\_7}

\bibitem{unpublished:tacconvex}
Cubuktepe, M., Jansen, N., Junges, S., Katoen, J., Topcu, U.: Convex
  optimization for parameter synthesis in {MDP}s. {IEEE} Trans. Autom. Control.
   (2022). \doi{10.1109/TAC.2021.3133265}

\bibitem{TVI}
Dai, P., Mausam, Weld, D.S., Goldsmith, J.: Topological value iteration
  algorithms. J. Artif. Intell. Res.  \textbf{42},  181--209 (2011),
  \url{https://www.jair.org/index.php/jair/article/view/10725}

\bibitem{DBLP:conf/cav/DutertreM06}
Dutertre, B., de~Moura, L.M.: A fast linear-arithmetic solver for {DPLL(T)}.
  In: {CAV}. {LNCS}, vol.~4144, pp. 81--94. Springer (2006)

\bibitem{DBLP:conf/icalp/Fearnley10}
Fearnley, J.: Exponential lower bounds for policy iteration. In: {ICALP} {(2)}.
  {LNCS}, vol.~6199, pp. 551--562. Springer (2010)

\bibitem{DBLP:conf/atva/ForejtKP12}
Forejt, V., Kwiatkowska, M.Z., Parker, D.: Pareto curves for probabilistic
  model checking. In: {ATVA}. {LNCS}, vol.~7561, pp. 317--332. Springer (2012).
  \doi{10.1007/978-3-642-33386-6\_25},
  \url{https://doi.org/10.1007/978-3-642-33386-6\_25}

\bibitem{DBLP:conf/tacas/FunkeJB20}
Funke, F., Jantsch, S., Baier, C.: Farkas certificates and minimal witnesses
  for probabilistic reachability constraints. In: {TACAS} {(1)}. {LNCS}, vol.
  12078, pp. 324--345. Springer (2020)

\bibitem{Gir14}
Giro, S.: Optimal schedulers vs optimal bases: An approach for efficient exact
  solving of {M}arkov decision processes. Theor. Comput. Sci.  \textbf{538},
  70--83 (2014). \doi{10.1016/j.tcs.2013.08.020},
  \url{https://doi.org/10.1016/j.tcs.2013.08.020}

\bibitem{DBLP:conf/issac/GleixnerSW12}
Gleixner, A.M., Steffy, D.E., Wolter, K.: Improving the accuracy of linear
  programming solvers with iterative refinement. In: {ISSAC}. pp. 187--194.
  {ACM} (2012)

\bibitem{soplex}
Gleixner, A.M., Steffy, D.E., Wolter, K.: Iterative refinement for linear
  programming. Tech. Rep.~3, ZIB, Takustr. 7, 14195 Berlin (2016).
  \doi{10.1287/ijoc.2016.0692}, \url{https://doi.org/10.1287/ijoc.2016.0692}

\bibitem{gurobi}
{Gurobi Optimization, LLC}: {Gurobi Optimizer Reference Manual} (2022),
  \url{https://www.gurobi.com}

\bibitem{HM14}
Haddad, S., Monmege, B.: Reachability in {MDP}s: Refining convergence of value
  iteration. In: {RP}. {LNCS}, vol.~8762, pp. 125--137. Springer (2014)

\bibitem{HM18}
Haddad, S., Monmege, B.: Interval iteration algorithm for {MDP}s and {IMDP}s.
  Theor. Comput. Sci.  \textbf{735},  111--131 (2018).
  \doi{10.1016/j.tcs.2016.12.003},
  \url{https://doi.org/10.1016/j.tcs.2016.12.003}

\bibitem{H22}
Hartmanns, A.: Correct probabilistic model checking with floating-point
  arithmetic. In: {TACAS} {(2)}. {LNCS}, vol. 13244, pp. 41--59. Springer
  (2022). \doi{10.1007/978-3-030-99527-0\_3},
  \url{https://doi.org/10.1007/978-3-030-99527-0\_3}

\bibitem{HH14}
Hartmanns, A., Hermanns, H.: The {M}odest {T}oolset: An integrated environment
  for quantitative modelling and verification. In: {TACAS}. {LNCS}, vol.~8413,
  pp. 593--598. Springer (2014). \doi{10.1007/978-3-642-54862-8\_51}

\bibitem{HK20-ovi}
Hartmanns, A., Kaminski, B.L.: Optimistic value iteration. In: {CAV} {(2)}.
  {LNCS}, vol. 12225, pp. 488--511. Springer (2020).
  \doi{10.1007/978-3-030-53291-8\_26},
  \url{https://doi.org/10.1007/978-3-030-53291-8\_26}

\bibitem{HKPQR19}
Hartmanns, A., Klauck, M., Parker, D., Quatmann, T., Ruijters, E.: The
  quantitative verification benchmark set. In: {TACAS}. {LNCS}, vol. 11427, pp.
  344--350. Springer (2019). \doi{10.1007/978-3-030-17462-0\_20}

\bibitem{HJKQV22}
Hensel, C., Junges, S., Katoen, J.P., Quatmann, T., Volk, M.: The probabilistic
  model checker {S}torm. Int. J. Softw. Tools Technol. Transf.  \textbf{24}(4),
   589--610 (2022). \doi{10.1007/s10009-021-00633-z}

\bibitem{DBLP:conf/cav/JungesTS20}
Junges, S., Torfah, H., Seshia, S.A.: Runtime monitors for {M}arkov decision
  processes. In: {CAV} {(2)}. {LNCS}, vol. 12760, pp. 553--576. Springer (2021)

\bibitem{DBLP:conf/cav/KelmendiKKW18}
Kelmendi, E., Kr{\"{a}}mer, J., Kret{\'{\i}}nsk{\'{y}}, J., Weininger, M.:
  Value iteration for simple stochastic games: Stopping criterion and learning
  algorithm. In: {CAV} {(1)}. {LNCS}, vol. 10981, pp. 623--642. Springer (2018)

\bibitem{KRSW22}
Kretinsky, J., Ramneantu, E., Slivinskiy, A., Weininger, M.: Comparison of
  algorithms for simple stochastic games. Inf. Comput.  (2022),
  \url{https://doi.org/10.1016/j.ic.2022.104885}

\bibitem{kumar2015history}
Kumar, A., Zilberstein, S.: History-based controller design and optimization
  for partially observable {MDP}s. In: ICAPS. vol.~25, pp. 156--164 (2015)

\bibitem{KNPS06}
Kwiatkowska, M.Z., Norman, G., Parker, D., Sproston, J.: Performance analysis
  of probabilistic timed automata using digital clocks. Formal Methods Syst.
  Des.  \textbf{29}(1),  33--78 (2006). \doi{10.1007/s10703-006-0005-2}

\bibitem{DBLP:conf/uai/LittmanDK95}
Littman, M.L., Dean, T.L., Kaelbling, L.P.: On the complexity of solving
  {M}arkov decision problems. In: {UAI}. pp. 394--402. Morgan Kaufmann (1995)

\bibitem{DBLP:journals/fmsd/MathurBCSV20}
Mathur, U., Bauer, M.S., Chadha, R., Sistla, A.P., Viswanathan, M.: Exact
  quantitative probabilistic model checking through rational search. Formal
  Methods Syst. Des.  \textbf{56}(1),  90--126 (2020)

\bibitem{MLG05}
McMahan, H.B., Likhachev, M., Gordon, G.J.: Bounded real-time dynamic
  programming: {RTDP} with monotone upper bounds and performance guarantees.
  In: {ICML}. {ACM} International Conference Proceeding Series, vol.~119, pp.
  569--576. {ACM} (2005). \doi{10.1145/1102351.1102423}

\bibitem{z3}
de~Moura, L.M., Bj{\o}rner, N.S.: {Z3:} an efficient {SMT} solver. In: {TACAS}.
  {LNCS}, vol.~4963, pp. 337--340. Springer (2008).
  \doi{10.1007/978-3-540-78800-3\_24},
  \url{https://doi.org/10.1007/978-3-540-78800-3\_24}

\bibitem{DBLP:conf/cav/PhalakarnTHH20}
Phalakarn, K., Takisaka, T., Haas, T., Hasuo, I.: Widest paths and global
  propagation in bounded value iteration for stochastic games. In: {CAV} {(2)}.
  {LNCS}, vol. 12225, pp. 349--371. Springer (2020)

\bibitem{puterman}
Puterman, M.L.: {M}arkov Decision Processes: Discrete Stochastic Dynamic
  Programming. Wiley Series in Probability and Statistics, Wiley (1994).
  \doi{10.1002/9780470316887}, \url{https://doi.org/10.1002/9780470316887}

\bibitem{QK18-svi}
Quatmann, T., Katoen, J.: Sound value iteration. In: {CAV} {(1)}. {LNCS}, vol.
  10981, pp. 643--661. Springer (2018). \doi{10.1007/978-3-319-96145-3\_37}

\bibitem{Saad1986GMRESAG}
Saad, Y., Schultz, M.H.: Gmres: a generalized minimal residual algorithm for
  solving nonsymmetric linear systems. Siam Journal on Scientific and
  Statistical Computing  \textbf{7},  856--869 (1986),
  \url{https://epubs.siam.org/doi/10.1137/0907058}

\bibitem{DBLP:journals/corr/abs-1305-5055}
Wimmer, R., Jansen, N., Vorpahl, A., {\'{A}}brah{\'{a}}m, E., Katoen, J.,
  Becker, B.: High-level counterexamples for probabilistic automata. Log.
  Methods Comput. Sci.  \textbf{11}(1) (2015)

\bibitem{WKHB08}
Wimmer, R., Kortus, A., Herbstritt, M., Becker, B.: Probabilistic model
  checking and reliability of results. In: {DDECS}. pp. 207--212. {IEEE}
  Computer Society (2008). \doi{10.1109/DDECS.2008.4538787}

\bibitem{PolPI}
Ye, Y.: The simplex and policy-iteration methods are strongly polynomial for
  the {M}arkov decision problem with a fixed discount rate. Mathematics of
  Operations Research  \textbf{36}(4),  593--603 (2011)

\end{thebibliography}
